\documentstyle[12pt,epsfig,amsfonts,amssymb,amsbsy]{article}
\def\beqa{\begin{eqnarray}}
\def\eeqa{\end{eqnarray}}
\def\nn{\nonumber}

\newcommand\VV{\setbox0=\hbox{V}\hbox{\rm V\raise\ht0
  \hbox to0pt{\hss\vbox to0pt{\hbox{v}\vss}}}}
\def\slashchar#1{\setbox0=\hbox{$#1$}           
   \dimen0=\wd0                                 
   \setbox1=\hbox{/} \dimen1=\wd1               
   \ifdim\dimen0>\dimen1                        
      \rlap{\hbox to \dimen0{\hfil/\hfil}}      
      #1                                        
   \else                                        
      \rlap{\hbox to \dimen1{\hfil$#1$\hfil}}   
      /                                         
   \fi}                                         %

\begin{document}

\begin{center}
{\large \bf Exclusive decays of $\Xi_{QQ'}^{\diamond}$ baryons
in NRQCD sum rules}\\ \vspace*{5mm} Andrei I. Onishchenko

\end{center}
\begin{center}
{\it Institute for Theoretical and Experimental Physics,\\ B.Cheremushkinskaja, 25, Moscow,
117259 Russia}.
\end{center}

\begin{abstract}
{We perform a detailed study of semileptonic form-factors for the
doubly heavy baryons in the framework of three-point NRQCD sum
rules. The analysis of spin symmetry relations as well as
numerical results on various exclusive decay modes of doubly heavy
baryons are given.}
\end{abstract}

\section{Introduction}

Besides an experimental search for an explanation of the
phenomenon of electroweak symmetry breaking and new physics beyond
the Standard Model, high energies and luminosities of current and
future particle accelerators provide a possibility to observe rare
processes with heavy quarks. An interesting topic here is a study
of physics of doubly heavy baryons. An analysis of dynamics of
these baryons can play a fundamental role in an extraction of
primary parameters of weak quark interactions. Due to distinctions
between the QCD effects inside the doubly and singly heavy
hadrons, one may strictly constrain incalculable nonperturbative
quantities, entering different schemes of calculations.

The real possibility of such experimental measurements was
recently confirmed by CDF Collaboration by the first observation
of $B_c$ meson \cite{cdf-bc}. As predicted theoretically
\cite{bc-rev}, this long-lived state of $\bar b$ and $c$ quarks
has the production cross sections, mass and decay rates, which are
compatible with the characteristic values for the doubly heavy
hadrons. Thus, the experimental search for the doubly heavy
baryons can also be successful. Of course, such the search would
be more strongly motivated if it would be supported by modern
theoretical studies and evaluations of basic characteristics for
the doubly heavy baryons.

Some steps forward this program were already done. First, the
production cross sections of doubly heavy baryons in hadron
collisions at high energies of colliders and in fixed target
experiments were calculated in the framework of perturbative QCD
for the hard processes and factorization of soft term related to
the nonperturbative binding of heavy quarks \cite{prod}. Second,
the lifetimes and branching fractions of some inclusive decay
modes were evaluated in the Operator Product Expansion combined
with the effective theory of heavy quarks, which results in series
over the inverse heavy quark masses and relative velocities of
heavy quarks inside the doubly heavy diquark \cite{ltime,DHD}.
Third, the families of doubly heavy baryons, which contain a set
of narrow excited levels in addition to the ground state, were
described in the framework of potential models \cite{pot}. The
picture of spectra, obtained in this analysis, is very similar to
that of heavy quarkonia. Fourth, the QCD and NRQCD sum rules
\cite{SVZ} were explored for the two-point baryonic currents in
order to calculate the masses and couplings of doubly heavy
baryons \cite{QCDsr,DHSR1,DHSR2}. And fifth, there are papers,
where exclusive semileptonic and some nonleptonic decay modes of
doubly heavy baryons in the framework of potential models and
within the Bethe-Salpeter approach were analyzed
\cite{Lozano,Guo}.

In the present paper we estimate form-factors for the semileptonic
decays of doubly heavy baryons together with semileptonic and some
nonleptonic decay modes in the framework of three-point NRQCD sum
rules for the case of spin $1/2$ - spin $1/2$ - baryon transitions
only. The estimates of contributions due to spin $1/2$-spin
$3/2$-baryon transitions are also given the basis of QCD
superflavor symmetry, emerging in the limit of very large heavy
quark masses. In the limit of zero recoil we derive the spin
symmetry relations on the form-factors, governing the semileptonic
transitions of doubly heavy baryons. The use of these symmetry
relations greatly simplifies  further evaluation of form-factors
for doubly heavy baryon transitions. A detailed analysis of baryon
couplings, needed to model the phenomenological part of
three-point sum rules is also provided. Our further exposition of
the obtained results is organized as follows. In Section 2 we
discuss our choice of interpolating currents between vacuum and
corresponding doubly heavy baryon state and give numerical
estimates of baryonic couplings in the framework of two-point
NRQCD sum rules. Section 3 is devoted to the description of the
method used to calculate the form-factors of interest. Here we
present a derivation of spin symmetry relations between
form-factors  in the limit of maximal invariant mass of leptonic
pair and give analytical expressions for corresponding double
spectral densities. In section 4 we present numerical estimates of
the form-factors studied together with predictions for the
semileptonic and some nonleptonic decay modes. And finally,
section 5 contains our conclusion.

\section{Two point sum rules}

In this section we describe  steps, required for the evaluation of
baryonic constants, as they will be needed later to model the
phenomenological part of three-point sum rules. The question,
which should be solved first is the choice of the corresponding
interpolating currents for the baryons under consideration. So, in
the next subsection we discuss their various choices and comment
on the merits of the prescription, used in this paper.

\subsection{Baryonic currents}

As was mentioned in Introduction, in this work we consider only
spin $1/2$ - spin $1/2$ transitions of doubly heavy baryons.
Hence, the discussion of baryonic currents later in this
subsection will be restricted to this case. For baryons,
containing two heavy quarks, there are two types of interpolating
currents:

\noindent 1) The prescription with the explicit spin structure of
the heavy diquark from the very beginning is given by
\begin{eqnarray}
J_{\Xi^{\prime \diamond}_{QQ^{\prime}}} &=& [Q^{iT}C\tau\gamma_5
Q^{j\prime}]q^k\varepsilon_{ijk},\nonumber\\
J_{\Xi_{QQ}^{\diamond}} &=& [Q^{iT}C\tau\boldsymbol{\gamma}^m
Q^j]\cdot\boldsymbol{\gamma}_m\gamma_5 q^k\varepsilon_{ijk},
\label{def}
\end{eqnarray}
Here $C$ is the charge conjugation matrix with the properties
$C\gamma_{\mu}^TC^{-1} = -\gamma_{\mu}$ and $C\gamma_5^TC^{-1} =
\gamma_5$, $i,j,k$ are color indices and $\tau $ is a matrix in
the flavor space.

\noindent 2) The currents, which require a further symmetrization
of heavy diquark wave function, have the form
\begin{eqnarray}
J_{\Xi_{QQ'}^{\diamond}} = \varepsilon^{\alpha\beta\gamma }
:(Q^{T}_{\alpha}C\gamma_5 q_{\beta })Q^{'}_{\gamma }:
\end{eqnarray}
The currents of the second type can be easily related to those of
first type by the Fierz rearrangement of quark fields and the
further symmetrization or antisymmetrization of diquark
wave-function, depending on the diquark spin state. The first type
of these currents was considered in \cite{DHSR1,DHSR2}, while the
currents of second type were discussed in \cite{QCDsr}. The
evaluation of form-factors, describing semileptonic decays of
doubly heavy baryons, what is the goal of this paper, is much easy
if we use the currents of the second type and make the
symmetrization or antisymmetrization of diquark wave-function at
the end of calculation on the level of form-factors. This
procedure will become clear in the section with our numerical
results for the exclusive decay modes of doubly heavy baryons. The
reasons here are the manifest symmetry relations for form-factors,
discussed by us later and rather simple calculations of various
spectral densities for three-point correlation functions, which
can also be done in full QCD framework without performing
complicated angular integrations. However, as was shown by the
authors of \cite{QCDsr}, for the second type of currents it is
difficult, in general, to achieve a stability of sum rules
predictions for the both extracted mass and coupling of doubly
heavy baryons. To the same time, all these difficulties are absent
for the currents of first type. So, the conclusion, which one may
do here, is that, the appropriate choice of baryonic currents
depends on the problem, which you would like to solve.

Thus, in view of the lack of experimental information on the
masses of doubly heavy baryons we will use for them the results of
two-point NRQCD sum rules for the first type of currents
\cite{DHSR1,DHSR2}. Taking these mass values as input, we
calculate further the baryonic couplings of the second type.

Next, let us discuss the couplings of strange heavy baryons, as
they are appear in the semileptonic decays of some doubly heavy
baryons. The currents, describing these hadrons, also classified
according to the symmetry properties of light diquark
wave-function. There are two spin $1/2$ $\Lambda $-type
(antisymmetric in the $q$ and $s$ - quarks\footnote{Here $q$
denotes one of the light quarks $u$ or $d$.}) and two $\Sigma
$-type (symmetric in the $q$ and $s$ - quarks) HQET currents,
namely
\begin{eqnarray}
J_{\Lambda 1} &=& (q^{T}C\gamma_5 s)Q_v,\quad J_{\Lambda 2} =
(q^{T}C\gamma^0\gamma_5 s)Q_v, \\ &&\quad\quad\quad
\mbox{($\Lambda$ - type)} \nonumber \\ J_{\Sigma 1} &=&
(q^{T}C\gamma^ks)\gamma^k\gamma_5Q_v,\quad J_{\Sigma 2} =
(q^{T}C\gamma^0\gamma^ks)\gamma^k\gamma_5Q_v, \\ &&\quad\quad\quad
\mbox{($\Sigma$ - type)} \nonumber
\end{eqnarray}
where $Q_v$ is the HQET heavy quark spinor, moving with velocity
$v$. The baryons, described by these currents, belong to the same
SU(4) multiplet, that have as its lowest level the $J^P =
\frac{1}{2}^{+}$ SU(3) octet. As, these hadrons contain only one
heavy quark, they belong to the second level of the mentioned
SU(4) multiplet. This level splits apart into two SU(3)
multiplets, a $\bf\bar 3$, states of which $\Xi_Q$ are
antisymmetric under interchange of two light quarks and thus
described by $\Lambda $ - type currents, and $\bf 6$, states of
which $\Xi_Q'$ are symmetric under interchange of light quark and
described by $\Sigma $ - type currents. Actually, there may be
some mixing between the pure $\bf\bar 3$ and $\bf 6$ states to
form the physical $\Xi_Q$ and $\Xi_Q^{'}$ states\footnote{They
both have the same $I$, $J$ and $P$ quantum numbers.}. So, in what
follows we will not distinguish between $\Xi_Q$ and $\Xi_Q^{'}$ -
baryons and will exploit the fact that both states have
non-vanishing overlap with $\varepsilon^{\alpha\beta\gamma }
:(q^{T}_{\alpha}C\gamma_5 Q_{\beta })s_{\gamma }:$ current. In
other words, what we suppose to calculate is the semileptonic
branching ratio of some of doubly heavy baryons into both $\Xi_Q$
and $\Xi_Q^{'}$ - baryons.

Now, let us briefly describe the two-point NRQCD sum rules, used
for their evaluation.

\subsection{Description of the method}

We start from the correlator of two baryonic currents with the
half spin
\begin{equation}
\Pi(p^2) = i\int d^4x e^{ipx}\langle 0|T\{J(x),\bar
J(0)\}|0\rangle = \slashchar{p} F_1(p^2) + F_2(p^2),\label{2pcor}
\end{equation}
Performing the OPE expansion of this correlation function, we get
a series, different terms of which give us the contributions of
operators with various dimensions. So, for $F_i$ ($i,2$)
functions we have the following expressions
\begin{equation}
F_i(p^2) = F_i^{pert}(p^2) + F_i^{\bar qq}(p^2)\langle\bar
qq\rangle + F_i^{G^2}(p^2)\langle\frac{\alpha_s}{\pi }G^2\rangle +
F_i^{mix}(p^2)\langle\bar qGq\rangle + \ldots
\end{equation}
To obtain theoretical expressions for the Wilson coefficients,
standing in front of different operators, one typically uses the
dispersion relation
\begin{equation}
F_i^{\diamond }(t) = \frac{1}{\pi
}\int_0^{\infty}\frac{\rho_i^{\diamond }(w)dw}{w-t},
\end{equation}
where $t =  k\cdot v$, $p_{\mu} = k_{\mu}+(m_1+m_2)v_{\mu}$ and
$\rho_i^{\diamond }$ denotes the imaginary part in the physical
region of corresponding Wilson coefficients in NRQCD\footnote{Here
$m_1$ and $m_2$ are the heavy quark masses and $v_{\mu}$ is
four-velocity of the baryon under consideration.}. The calculation
of spectral densities $\rho_i^{\diamond }$ proceeds through the
use of Cutkosky rules \cite{Cutk} and for the case of
$\varepsilon^{\alpha\beta\gamma } :(Q^{T}_{\alpha}C\gamma_5
q_{\beta })Q^{'}_{\gamma }:$ current and different quark masses
was done in the QCD framework in \cite{QCDsr}. Here we use the
results of this work. The needed NRQCD spectral densities  were
obtained by simple NRQCD expansion of corresponding QCD
expressions and the results of this expansion could be found in
Appendix~A.

To relate the NRQCD correlators to hadrons, we use the dispersion
representation for the two-point function with the physical
spectral density, given by appropriate resonance and continuum
part. The coupling constants of doubly heavy baryons are defined
by the following expression
\begin{eqnarray}
\langle 0|J_H|H(p)\rangle = i Z_H u(v,M_H)e^{ip\cdot x},
\end{eqnarray}
where $p=M_H v$ and the spinor field with four-velocity $v$ and
mass $M_H$ satisfies the equation $\slashchar{v}u(v,M_H) =
u(v,M_H)$.

We suppose that the continuum part, starting from the threshold
value $w_{cont}$, is equal to that of calculated in the framework
of NRQCD. Then, equalizing the correlators, calculated in NRQCD
and given by the physical states, the integrations above
$w_{cont}$ cancel each other in both sides of sum rules relation.
Further, we write down the correlators at the deep under-threshold
point $t_0 = -(m_1+m_2)+t$ at $t\to 0$.

Introducing the following notation for the $n$-th moment of
two-point correlation function
\begin{equation}
{\cal M}_n^{i} = \frac{1}{\pi }\int_0^{w_{cont}}\frac{\rho_i
(w)dw}{(w+m_1+m_2)^{n}},
\end{equation}
and using the approximation of single bound state pole, we can
write the following relation
\begin{eqnarray}
{\cal M}_n^i = |Z_H^{[i]}|^2\frac{1}{M_H^{n+1-i}}.
\end{eqnarray}
From which one can read off the corresponding expression for the
baryon coupling in the moment scheme
\begin{equation}
|Z_H^{[i]}|^2 = {\cal M}_n^i M_H^{n+1-i},
\end{equation}
where we see the dependence of sum rules on the scheme parameter
$n$. Therefore, we will tend to find the region of parameter
values, where, first, the result is stable under the variation of
moment number $n$, and, second, the both correlators $F_1$ and
$F_2$ reproduce equal values of coupling constants. In the QCD sum
rule analysis of \cite{QCDsr} there was a problem, that values of
baryon coupling constants, obtained from $F_1$ and $F_2$
correlation functions significantly differ. To cure this problem,
in \cite{DHSR1} it was proposed to include in calculations also
contributions, coming from the OPE expansion for the correlator of
two quark fields \cite{Smilga}
\begin{equation}
\langle 0|T\{q_i^a(x)\bar q_j^b(0)\}|0\rangle =
-\frac{1}{12}\delta^{ab}\delta_{ij}\langle\bar qq\rangle\cdot
\left[1+\frac{m_0^2x^2}{16}+\frac{\pi^2x^4}{288}\langle\frac{\alpha_s}{\pi}G^2\rangle
+ ...\right].
\end{equation}
With an account of these corrections the quark condensate
contribution to moments gets modified
\begin{equation}
{\cal M}_n^{\bar qq} = {\cal M}_n^{\bar
qq}-\frac{(n+2)!}{n!}\frac{m_0^2}{16}{\cal M}_{n+2}^{\bar
qq}+\frac{(n+4)!}{n!}\frac{\pi^2}{288}\langle\frac{\alpha_s}{\pi}G^2\rangle
{\cal M}_{n+4}^{\bar qq}.
\end{equation}
The derivation of two-point HQET sum rules for the heavy baryons
with the strangeness follows the same lines as that for baryons
with two heavy quarks. Here we only comment on the differences. To
obtain the HQET expressions for the spectral densities  we again
use the QCD result of \cite{QCDsr}. However, the transition
between the QCD expressions for the doubly heavy baryons and HQET
expressions for the heavy baryons with the strangeness is more
intricate. First, we should take a limit when one of the heavy
quark masses goes to zero and second, we should allow this quark
to condense. We have calculated explicitly the $s$ -quark
condensate contribution to the both $F_1$ and $F_2$ correlation
functions and subtracted $1/m_s$ poles from gluon condensate
contribution, related to the strange quark condensate due to the
following heavy quark expansion
\begin{equation}
\langle\bar ss\rangle \Rightarrow
-\frac{1}{12m_s}\frac{\alpha_s}{\pi}\langle G^2\rangle -
\frac{1}{360m_s^3}\frac{\alpha_s}{\pi}\langle G^3\rangle + \ldots
\end{equation}
The appearing logarithmic singularities can be related to the
mixed quark condensate with the help of the same heavy quark
expansion
\begin{equation}
\langle\bar sGs\rangle \Rightarrow \frac{m_s}{2}\log
m_s^2\frac{\alpha_s}{\pi}\langle G^2\rangle -
\frac{1}{12m_s}\frac{\alpha_s}{\pi}\langle G^3\rangle + \ldots
\end{equation}
And, finally, one must subtract the nonsingular gluon condensate
contribution, belonging to the quark condensate, what can be
easily done with the calculated explicit expression for quark
condensate contribution. The resulted HQET spectral densities for
the case of ordinary baryons with strangeness were collected by us
in Appendix A. For the strange quark condensate contribution, as
in the case of light quark condensate, we also take into account
the corrections, coming from the OPE expansion  for the correlator
of two strange quark fields \cite{DHSR2}
\begin{eqnarray}
\langle 0|T\{s_i^a(x)\bar s_j^b(0)\}\rangle &=&
-\frac{1}{12}\delta^{ab}\delta_{ij}\langle\bar
ss\rangle\cdot\left[1+\frac{x^2(m_0^2-2m_s^2)}{16}+\frac{x^4(\pi^2\langle
\frac{\alpha_sG^2}{\pi}\rangle
-\frac{3}{2}m_s^2(m_0^2-m_s^2))}{288}\right] \nonumber \\ &+&
im_s\delta^{ab}x_{\mu }\gamma^{\mu}_{ij}\langle\bar
ss\rangle\left[\frac{1}{48}+\frac{x^2}{24^2}\left(\frac{3m_0^2}{4}-m_s^2\right)\right].
\end{eqnarray}
With this corrections the $s$-quark contribution to the moments
for the $F_1$ and $F_2$ correlation functions has the form
\begin{eqnarray}
{\cal M}_1^{\bar ss}(n) &=& -\frac{1}{4}m_s\frac{(n+1)!}{n!}{\cal
M}^{\bar
ss}(n+1)+\frac{1}{48}\left(\frac{3m_0^2}{4}-m_s^2\right)m_s\frac{(n+3)!}{n!}{\cal
M}^{\bar ss}(n+3), \\ {\cal M}_2^{\bar ss}(n) &=& {\cal M}^{\bar
ss}(n)-\frac{m_0^2-2m_s^2}{16}\frac{(n+2)!}{n!}{\cal M}^{\bar
ss}(n+2)+\nonumber \\ &&
\frac{(\pi^2\langle\frac{\alpha_sG^2}{\pi}\rangle
-\frac{3}{2}m_s^2(m_0^2-m_s^2))}{288}\frac{(n+4)!}{n!}{\cal
M}^{\bar ss}(n+4),
\end{eqnarray}
where
\begin{equation}
{\cal M}^{\bar ss}(n) = \frac{1}{\pi
}\int_0^{w_{cont}}\frac{\rho^{\bar ss} (w)dw}{(w+m_Q+m_s)^{n}},
\end{equation}
and
\begin{equation}
\rho^{\bar ss} (w) = -\frac{\langle\bar
ss\rangle(m_s+w)^2(2m_Q+m_s+w)^2}{4\pi (m_Q+m_s+w)}.
\end{equation}
Now, having all theoretical expressions for baryon coupling
constants in the moment scheme, we will proceed in the next
subsection with the numerical estimates.

\setlength{\unitlength}{1mm}
\begin{figure}[th]
\begin{center}
\vspace*{0.2cm}
\begin{picture}(200,150)
\put(5,20){\epsfxsize=7cm \epsfbox{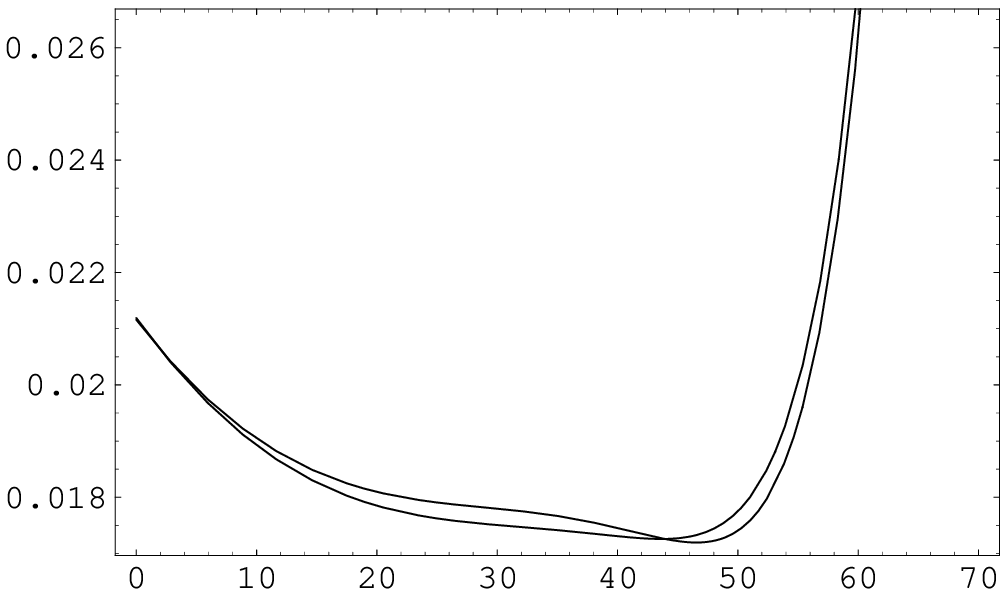}}
\put(95,20){\epsfxsize=7cm \epsfbox{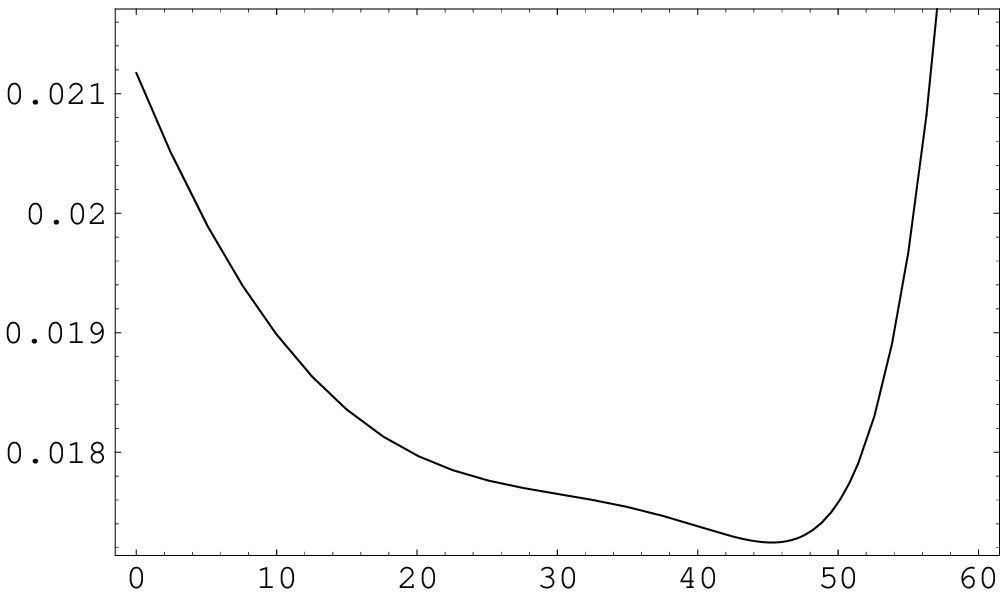}}
\put(0,155){$|Z^{[1,2]}_n|^2$, GeV$^3$}
\put(90,155){$\frac{|Z^{[1]}_n|^2+|Z^{[2]}_n|^2}{2}$, GeV$^3$}
\put(70,105){$n$} \put(160,105){$n$} \put(45,100){a)}
\put(135,100){b)}
\end{picture}
\end{center}
\vspace*{-10cm} \caption{a) The values of $|Z^{[1]}_n|^2$ and
$|Z^{[1]}_n|^2$ $\Xi_{bb}^{\diamond }$ - baryon couplings as
functions of the moment number $n$; b) the value of
$\frac{|Z^{[1]}_n|^2+|Z^{[2]}_n|^2}{2}$ average as function of the
moment number $n$ for the case of $\Xi_{bb}^{\diamond }$ -
baryons.} \label{2bb}
\end{figure}
\normalsize

\setlength{\unitlength}{1mm}
\begin{figure}[th]
\begin{center}
\begin{picture}(200,150)
\put(5,20){\epsfxsize=7cm \epsfbox{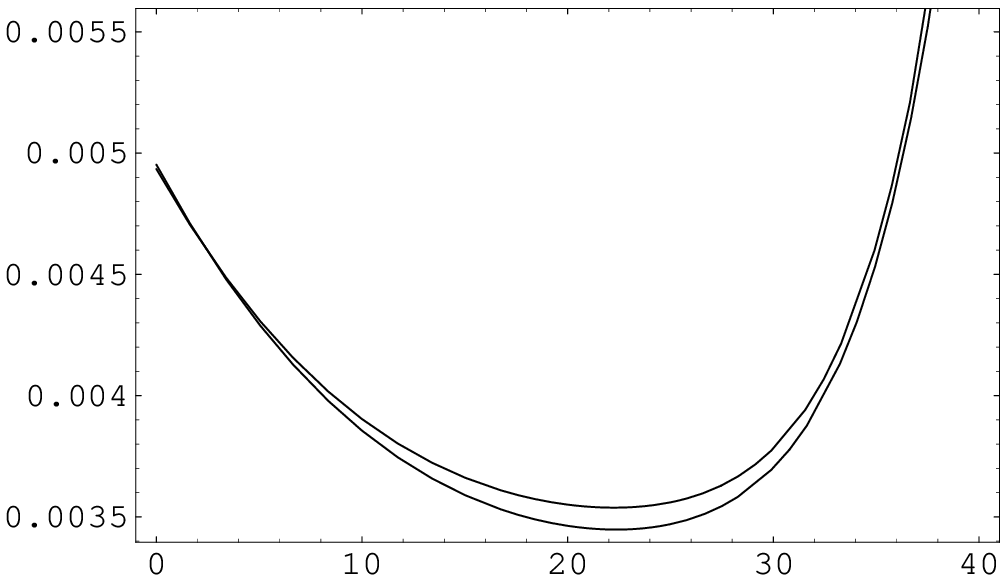}}
\put(95,20){\epsfxsize=7cm \epsfbox{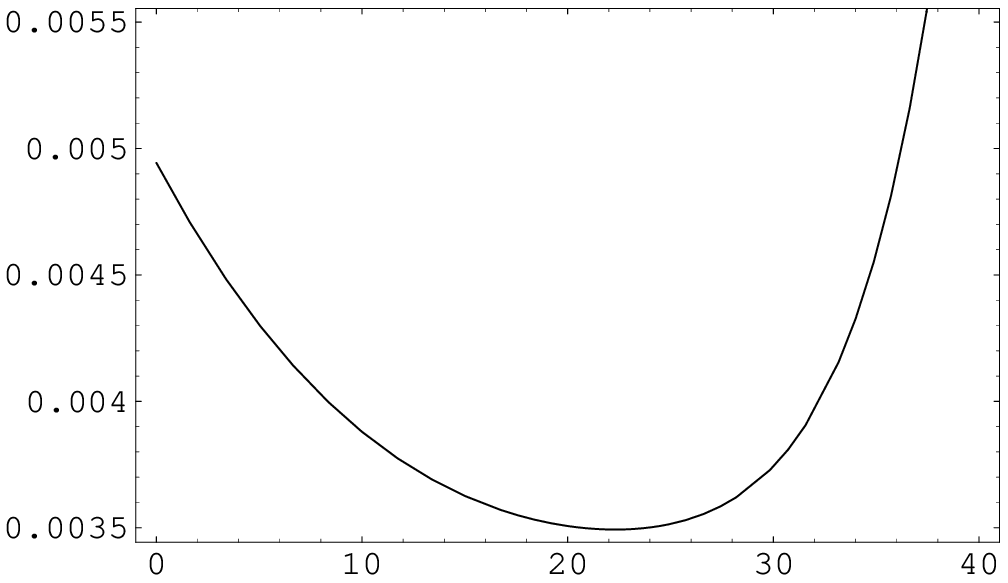}}
\put(0,155){$|Z^{[1,2]}_n|^2$, GeV$^3$}
\put(90,155){$\frac{|Z^{[1]}_n|^2+|Z^{[2]}_n|^2}{2}$, GeV$^3$}
\put(70,105){$n$} \put(160,105){$n$} \put(45,103){a)}
\put(135,103){b)}
\end{picture}
\end{center}
\vspace*{-11.cm} \caption{The results for the $\Xi_{bc}^{\diamond
}$ - baryon interpolating current in the case of decaying
$b$-quark (the $(c^TC\gamma_5q)b$-current ): a) the values of
$|Z^{[1]}_n|^2$ and $|Z^{[1]}_n|^2$ $\Xi_{bc}^{\diamond }$ -
baryon couplings as functions of the moment number $n$; b) the
value of $\frac{|Z^{[1]}_n|^2+|Z^{[2]}_n|^2}{2}$ average as
function of the moment number $n$ for the case of
$\Xi_{bc}^{\diamond }$ - baryons.} \label{2bc1}
\end{figure}
\normalsize
\setlength{\unitlength}{1mm}
\begin{figure}[th]
\begin{center}
\vspace*{1.4cm}
\begin{picture}(200,150)
\put(5,35){\epsfxsize=7cm \epsfbox{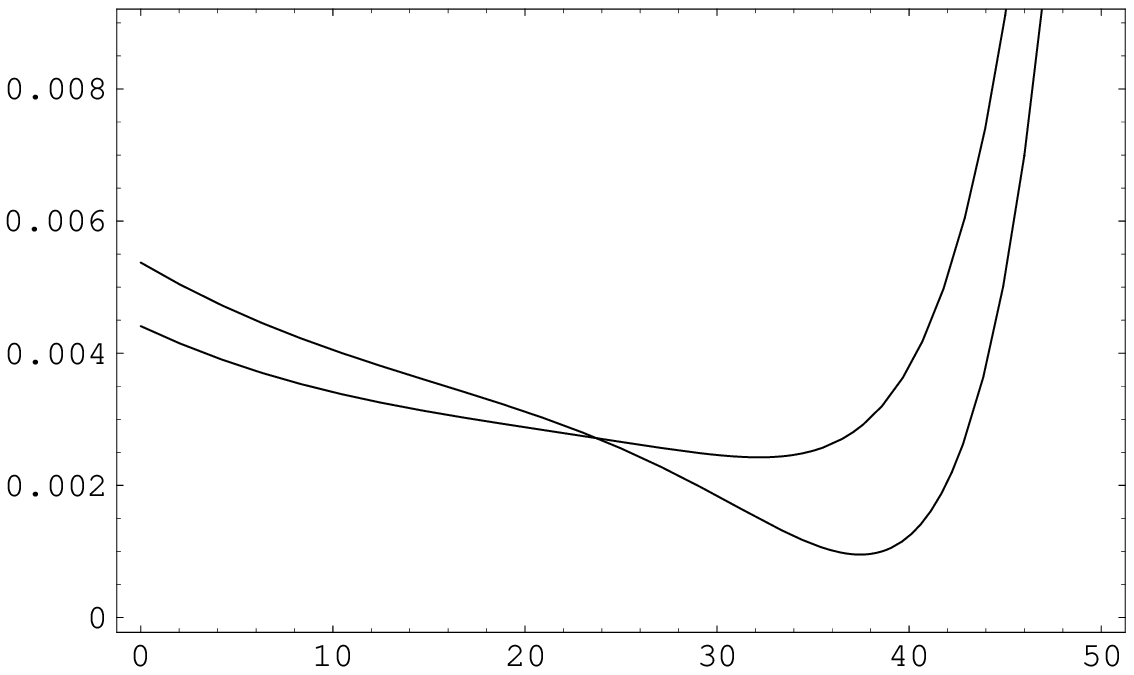}}
\put(95,35){\epsfxsize=7cm \epsfbox{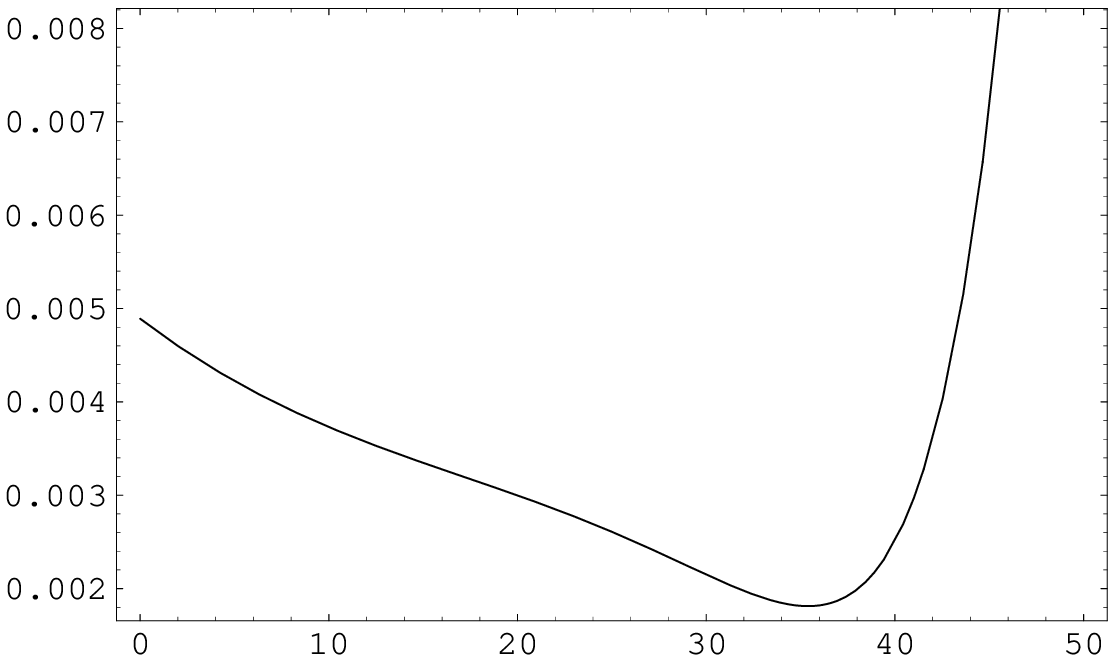}}
\put(0,168){$|Z^{[1,2]}_n|^2$, GeV$^3$}
\put(90,168){$\frac{|Z^{[1]}_n|^2+|Z^{[2]}_n|^2}{2}$, GeV$^3$}
\put(70,118){$n$} \put(160,118){$n$} \put(45,116){a)}
\put(135,116){b)}
\end{picture}
\end{center}
\vspace*{-12.4cm} \caption{The results for the $\Xi_{bc}^{\diamond
}$ - baryon interpolating current in the case of decaying
$c$-quark (the $(b^TC\gamma_5q)c$-current ): a) the values of
$|Z^{[1]}_n|^2$ and $|Z^{[1]}_n|^2$ $\Xi_{bc}^{\diamond }$ -
baryon couplings as functions of the moment number $n$; b) the
value of $\frac{|Z^{[1]}_n|^2+|Z^{[2]}_n|^2}{2}$ average as
function of the moment number $n$ for the case of
$\Xi_{bc}^{\diamond }$ - baryons.} \label{2bc2}
\end{figure}
\normalsize

\setlength{\unitlength}{1mm}
\begin{figure}[th]
\begin{center}
\begin{picture}(200,150)
\put(50,-5){\epsfxsize=8cm \epsfbox{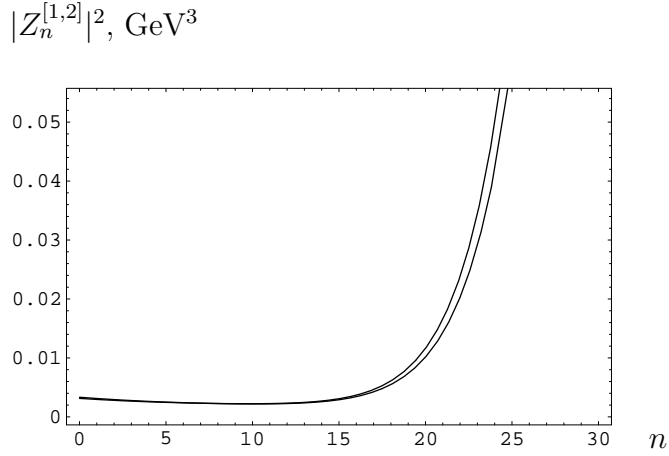}}
\put(50,150){$|Z^{[1,2]}_n|^2$, GeV$^3$} \put(135,95){$n$}
\end{picture}
\end{center}
\vspace*{-10cm} \caption{The values of $|Z^{[1]}_n|^2$ and
$|Z^{[2]}_n|^2$ $\Xi_{cc}^{\diamond }$ - baryon couplings as
functions of the moment number $n$.} \label{2cc}
\end{figure}
\normalsize

\setlength{\unitlength}{1mm}
\begin{figure}[th]
\begin{center}
\vspace*{0.5cm}
\begin{picture}(200,150)
\put(5,20){\epsfxsize=7cm \epsfbox{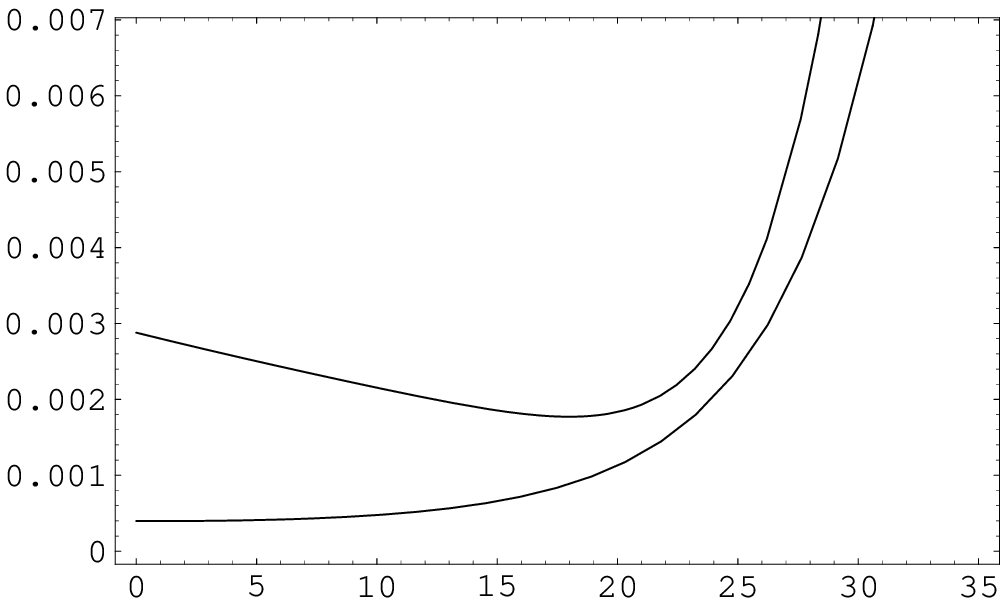}}
\put(95,20){\epsfxsize=7cm \epsfbox{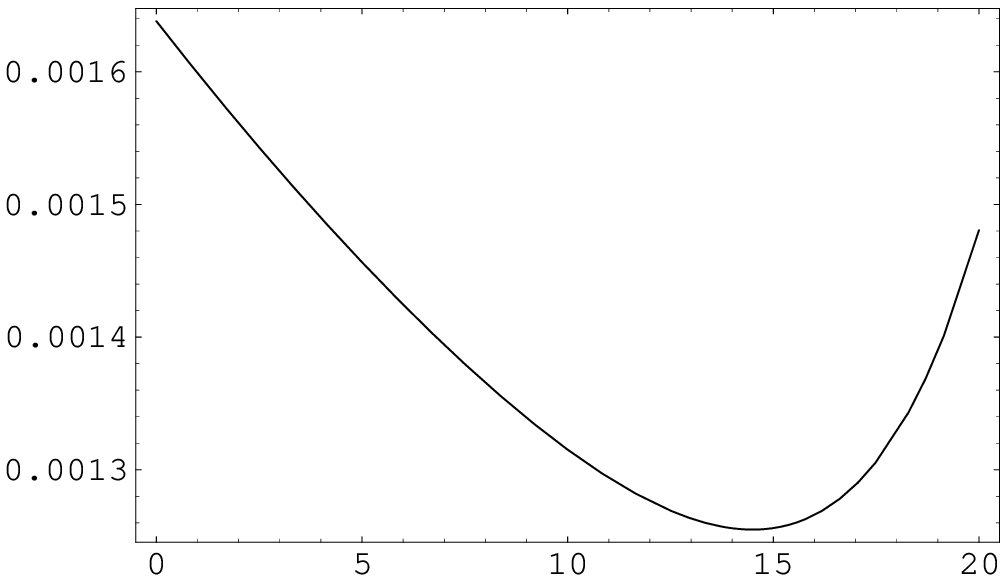}}
\put(0,155){$|Z^{[1,2]}_n|^2$, GeV$^3$}
\put(90,155){$\frac{|Z^{[1]}_n|^2+|Z^{[2]}_n|^2}{2}$, GeV$^3$}
\put(70,105){$n$} \put(160,105){$n$} \put(45,102){a)}
\put(135,102){b)}
\end{picture}
\end{center}
\vspace*{-10.7cm} \caption{a) The values of $|Z^{[1]}_n|^2$ and
$|Z^{[2]}_n|^2$ $\Xi_{b}^{\diamond }$ - baryon couplings as
functions of the moment number $n$; b) the value of
$\frac{|Z^{[1]}_n|^2+|Z^{[2]}_n|^2}{2}$ average as function of the
moment number $n$ for the case of $\Xi_{b}^{\diamond }$ -
baryons.} \label{2bs}
\end{figure}
\normalsize

\setlength{\unitlength}{1mm}
\begin{figure}[th]
\begin{center}
\vspace*{0.3cm}
\begin{picture}(200,150)
\put(5,20){\epsfxsize=7cm \epsfbox{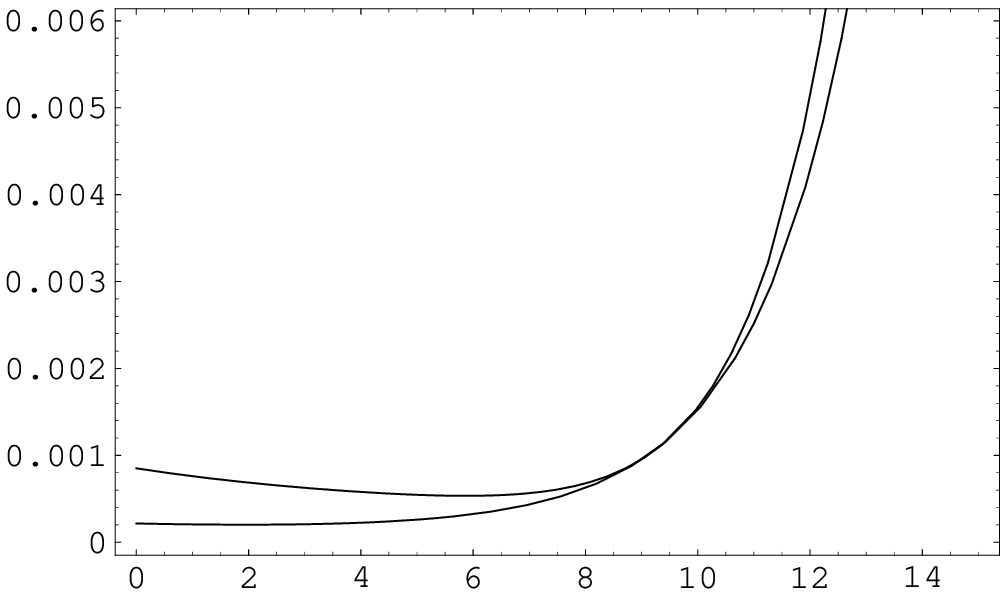}}
\put(95,20){\epsfxsize=7cm \epsfbox{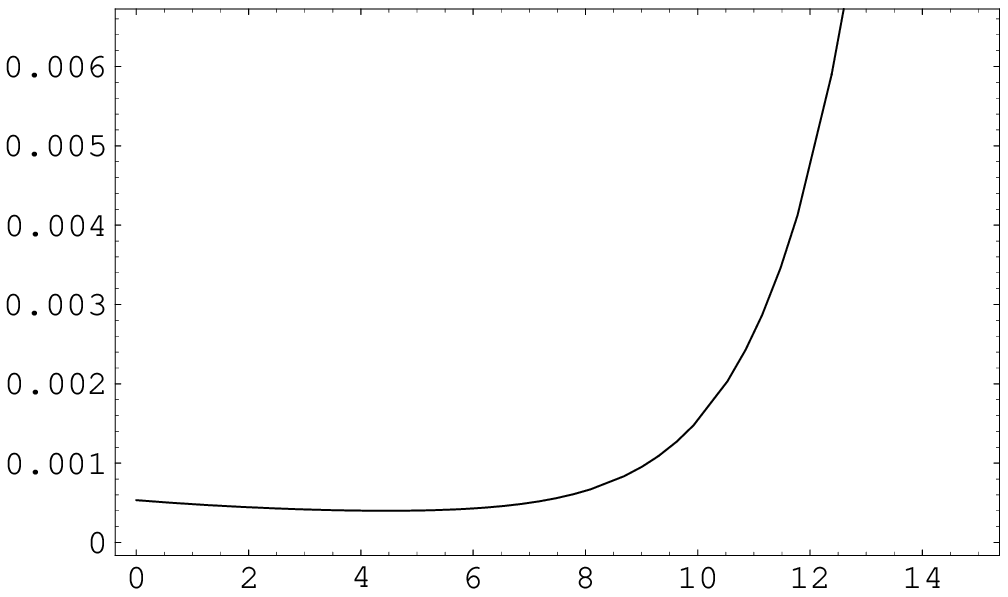}}
\put(0,155){$|Z^{[1,2]}_n|^2$, GeV$^3$}
\put(90,155){$\frac{|Z^{[1]}_n|^2+|Z^{[2]}_n|^2}{2}$, GeV$^3$}
\put(70,105){$n$} \put(160,105){$n$} \put(45,100){a)}
\put(135,100){b)}
\end{picture}
\end{center}
\vspace*{-10cm} \caption{a) The values of $|Z^{[1]}_n|^2$ and
$|Z^{[2]}_n|^2$ $\Xi_{c}^{\diamond }$ - baryon couplings as
functions of the moment number $n$; b) the value of
$\frac{|Z^{[1]}_n|^2+|Z^{[2]}_n|^2}{2}$ average as function of the
moment number $n$ for the case of $\Xi_{c}^{\diamond }$ -
baryons.} \label{2cs}
\end{figure}
\normalsize

\subsection{Numerical estimates}

In this subsection we present the results on the coupling
constants of doubly heavy baryons and the heavy baryons with the
strangeness. In the scheme of moments, which we employ here to
extract the baryonic couplings from the two-point sum rules, the
dominant uncertainty in estimates comes from the variation of
heavy quark mass values. In the analysis we chose the following
region of quark mass values
\begin{equation}
m_b = 4.6 - 4.7\mbox{~GeV},\quad m_c = 1.35 - 1.40\mbox{~GeV},
\end{equation}
what is the ordinary choice used in sum rules estimates of heavy
quarkonia. For the strange quark mass we use the value $m_s =
0.15$ GeV.

Next point, we would like to discuss, is an account of Coulomb
corrections inside the doubly heavy diquark. As is well known,
these corrections give large contribution to baryon coupling
constants and are essential for relative contributions of
perturbative and condensate terms to the correlator \cite{Valera}.
With a good accuracy at low or moderate values of moment number,
the effect of Coulomb interactions can be written as overall
Zommerfeld factor in front of perturbative spectral density of
heavy subsystem for the square of baryon coupling. But, as it was
shown in \cite{Valera}, the same Zommerfeld factors should be
taken into account in calculations of three-point correlation
functions considered later in this paper. It occurs, that for the
form-factors for the semileptonic transitions of doubly heavy
baryons these corrections cancel each other in average. So, the
calculation of desired form-factors for the doubly heavy baryons
can be consistently performed without accounting for Coulomb
corrections either, provided we neglect them both in the two-point
and three-point sum rules. This is the approach, we will follow in
the present work for the evaluation of form-factors.

The dependence of estimates on the threshold of continuum
contribution in the two-point sum rules is not so valuable as on
quark masses. We fix the region of $w_{cont}$ as
\begin{equation}
w_{cont} = 1.3 - 1.4\mbox{~GeV},
\end{equation}
which is in agreement with our previous estimates of doubly heavy
baryon coupling of the first type currents in the same framework
of two-point NRQCD sum rules. For the condensates of quark and
gluons we use the following regions:
\begin{equation}
\langle\bar qq\rangle = -(250-270\mbox{~MeV})^3,\quad m_0^2 =
0.75-0.85\mbox{~GeV}^2,\quad\langle\frac{\alpha_s}{\pi}G^2\rangle
= (1.5-2)\cdot 10^{-2}\mbox{~GeV}^4,
\end{equation}
and
\begin{equation}
\langle\bar ss\rangle = 0.8\pm 0.2~\langle\bar qq\rangle
\end{equation}
As we already mentioned, for the second type of currents used
here, we evaluate the coupling constants only and use the masses
of doubly heavy baryons, calculated by us previously \cite{DHSR1},
as inputs
\begin{equation}
M_{\Xi_{cc}} = 3.47\pm 0.05\mbox{~GeV},\quad M_{\Xi_{bc}} =
6.80\pm 0.05\mbox{~GeV},\quad M_{\Xi_{bb}} = 10.07\pm
0.09\mbox{~GeV},
\end{equation}
which are in agreement with the values obtained in the framework
of potential models. For the masses of heavy baryons with the
strangeness, appearing as products of semileptonic decays of some
of the doubly heavy baryons, we use the following values:
\begin{equation}
M_{\Xi_b} = 5.8\mbox{~GeV},\quad M_{\Xi_c} = 2.45\mbox{~GeV}.
\end{equation}
Figs. 1-6 show the dependence of baryon couplings on the momentum
number. We find that the stability regions for $|Z_{1(2)}|^2$
determined from the $F_1$ and $F_2$ correlators coincide with
those, obtained in the analysis of two-point correlation functions
for the first type of currents \cite{DHSR1}. However, for some of
the couplings, calculated here, we see a sizeable difference in
the predictions coming from the $F_1$ and $F_2$ correlation
functions. This problem could not be alleviated by the variation
of parameters. So, in order to determine the corresponding
coupling values to be used in the phenomenological part of
three-point sum rules, we consider an average coupling for these
currents, whose square is given by the average of squares for
$Z_1$ and $Z_2$ couplings. The resulted values of baryonic
coupling are
\begin{eqnarray}
&& |Z_{bb}|^2 = 1.7 ~10^{-2}\mbox{~GeV}^6,\quad |Z_{cc}|^2 = 2.3
~10^{-3}\mbox{~GeV}^6, \\ && |Z_{bc}^1|^2 = 3.5
~10^{-3}\mbox{~GeV}^6,\quad |Z_{bc}^2|^2 = 1.8
~10^{-3}\mbox{~GeV}^6,\\ && ~~
(c^TC\gamma_5q)b-\mbox{current}\quad ~~~~~
(b^TC\gamma_5q)c-\mbox{current}\nonumber \\ && |Z_{bs}|^2 = 1.3
~10^{-3}\mbox{~GeV}^6,\quad |Z_{cs}|^2 = 4.3
~10^{-4}\mbox{~GeV}^6.\nonumber
\end{eqnarray}
Having estimates for the couplings of initial and final state
baryons with respect to semileptonic transitions, we will continue
in the next section with the determination of form-factors.

\section{Three-point sum rules}

In this section we describe our framework for the calculation of
form-factors, governing the semileptonic decays of doubly heavy
baryons. Here we derive the spin symmetry relations between
various form-factors, arising in the limit of the maximal
invariant mass of leptonic pair and give analytical expressions
for corresponding spectral densities.

\setlength{\unitlength}{1mm}
\begin{figure}[th]
\begin{center}
\vspace*{2.cm}
\begin{picture}(200,150)
\put(50,-65){\epsfxsize=8cm \epsfbox{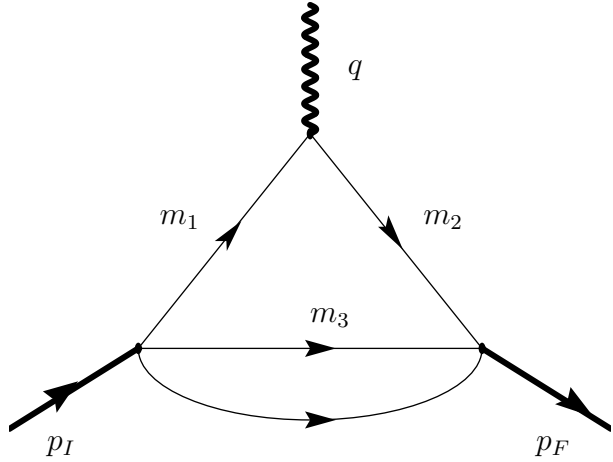}}
\put(70,140){$m_1$} \put(105,140){$m_2$} \put(90,127){$m_3$}
\put(55,110){$p_I$} \put(120,110){$p_F$} \put(95,160){$q$}
\end{picture}
\end{center}
\vspace*{-11.5cm} \caption{The diagram, corresponding to the
three-point correlation function considered in the paper.}
\label{diagram}
\end{figure}
\normalsize

Following the standard procedure for the evaluation of
form-factors in the framework of QCD sum rules, we consider the
three-point function
\begin{eqnarray}
\Pi_{\mu } = i^2\int d^4xd^4y \langle
0|T\{J_{H_F}(x)J_{\mu}(0)\bar J_{H_I} \}|0\rangle e^{i p_F\cdot
x}e^{-i p_I\cdot y},
\end{eqnarray}
where $J_{\mu}$ is the vector or axial transition current, matrix
elements of which between baryonic ground states we would like to
calculate. Fig.7 shows a diagram, corresponding to the mentioned
three-point function. The theoretical expression for the
three-point correlation function can be easily calculated with the
use of double dispertion relation
\begin{eqnarray}
\Pi_{\mu }^{(theor)}(s_1,s_2,q^2) = \frac{1}{(2\pi
)^2}\int_{m_I^2}^{\infty}ds_1\int_{m_F^2}^{\infty}ds_2\frac{\rho_{\mu
}(s_1,s_2,q^2)}{(s_1-s_1^0)(s_2-s_2^0)} + \mbox{subtractions},
\end{eqnarray}
where the desired spectral density could be obtained with the help
of Cutkosky rules \cite{Cutk}. We will continue with the
calculation of spectral  densities later in this paper after
discussing the spin symmetry relations for form-factors. The
latter, as will be seen, greatly simplify the calculations to be
done. Now, let us discuss the phenomenological part of three point
sum rules under consideration.  Saturating the channels of initial
and final state hadrons by ground states of corresponding baryons,
we have the following phenomenological expression for the
three-point correlation function:
\begin{eqnarray}
\Pi_{\mu }^{(phen)}(s_1,s_2,q^2) &=& \sum_{spins}\frac{\langle
0|J_{H_F}|H_F(p_F)\rangle}{s_2^0-M_{H_F}^2}\times \\ && \langle
H_F(p_F)|J_{\mu}|H_I(p_I)\rangle\frac{\langle H_I (p_I)|\bar
J_{H_I}|0\rangle }{s_1^0-M_{H_I}^2} \nonumber
\end{eqnarray}
The formfactors for the weak spin $\frac{1}{2}$ -- spin
$\frac{1}{2}$ baryon transitions are usually modeled as following:
\begin{eqnarray}
\langle H_F(p_F)|J_\mu|H_I(p_I)\rangle&=& \\
&&\hspace{-3cm}\overline{u}(p_F) [\gamma_\mu
(F_1^V+F_1^A\gamma_5)+i\sigma_{\mu\nu}q^\nu(F_2^V+F_2^A\gamma_5)+
q_\mu (F_3^V+F_3^A\gamma_5)]u(p_I)\; \nonumber
\end{eqnarray}
However, in the NRQCD limit it is  more convenient to use an alternative
parametrization
\begin{eqnarray}
\langle H_F(p_F)|J_{\mu}|H_I(p_I)\rangle &=& \\ &&\hspace{-3cm}
\bar u(p_F) \left( \gamma_{\mu} G^V_1 + v_{\mu }^I G^V_2 + v_{\mu
}^F G^V_3 +\gamma_5 (\gamma_{\mu} G^A_1 + v_{\mu }^I G^A_2 +
v_{\mu }^F G^A_3 )\right) u(p_I)\nonumber,
\end{eqnarray}
where these  two parametrizations can be related to each other
with the help of the following relations: \beqa F^V_1(t) &=& G^V_1
+(m_F+m_I)\left(\frac{1}{2 m_I} G^V_2+\frac{1}{2 m_F} G^V_3\right)
~ ,  \nn\\ F^V_2(t) &=& -\frac{1}{2 m_I} G^V_2-\frac{1}{2 m_F}
G^V_3 ~ , \nn \\ F^V_3(t) &=& -\frac{1}{2 m_I} G^V_2+\frac{1}{2
m_F} G^V_3  ~ ,\nn \\ F^A_1(t) &=& -G^A_1
-(m_F-m_I)\left(\frac{1}{2 m_I} G^A_2+\frac{1}{2 m_F} G^A_3\nn
\right) ~ ,  \\ F^A_2(t) &=& \frac{1}{2 m_I} G^A_2+\frac{1}{2 m_F}
G^A_3 ~ ,  \nn \\ F^A_3(t) &=& \frac{1}{2 m_I} G^A_2-\frac{1}{2
m_F} G^A_3 ~ . \label{FG} \eeqa Naively, all these six formfactors
in either parametrization are independent, but, as we will show in
the next subsection, in the limit of zero recoil the semileptonic
decays of doubly heavy baryons can be described by the only
universal function, an analogue of Isgur-Wise function.

\subsection{\large\bf Symmetry relations}

Now, let us discuss the spin symmetry relations among the
form-factors, arising in the limit of zero recoil for the final
state baryon. That is, we consider a limit\footnote{For the
discussion of this limit see \cite{Valera}.}, where $v_I\neq v_F$
and $\omega = (v_I\cdot v_F) \to 1$. The theoretical expression
for the three-point correlation function for the case of heavy to
heavy underlying quark transition in this limit has the following
form:
\begin{equation}
\Pi_{\mu}^{(theor)} \sim
\xi^{IW}(q^2)(1+\slashchar{\tilde{v}}_F)\gamma_{\mu}(1-\gamma_5)
(1+\slashchar{\tilde{v}}_I),
\end{equation}
where
\begin{eqnarray}
\tilde{v}_I &=& v_I + \frac{m_3}{2m_1}(v_I-v_F) \\ \tilde{v}_F &=&
v_F + \frac{m_3}{2m_2}(v_F-v_I).
\end{eqnarray}
So, for this type of transitions we have already, from the very
beginning, the only universal function and no further analysis is
required. The theoretical expression for the three-point
correlation function in the case of heavy to light underlying
quark transition has more complicated form
\begin{equation}
\Pi_{\mu}^{(theor)} \sim \{\xi_1(q^2)\slashchar{v}_I +
\xi_2(q^2)\slashchar{v}_F + \xi_3(q^2)\}\gamma_{\mu}(1-\gamma_5)
(1+\slashchar{\tilde{v}}_I)
\end{equation}
Considering different convolutions of the theoretical and
phenomenological three-point correlation functions with Lorenz
structures made of hadron velocities and $\gamma $ - matrices and
equating them, we obtain two relations on the semileptonic
form-factors in this case
\begin{eqnarray}
(G_1^V+G_2^V+G_3^V) &=& \xi^{IW}(q^2) \\ G_1^A &=& \xi^{IW}(q^2)
\end{eqnarray}
and a relation between $\xi_i(q^2)$ functions $$ \xi_1(q^2) +
\xi_2(q^2) = \xi_3(q^2) = \xi^{IW}(q^2) $$ Recalling also, that in
any considered transition we always have heavy baryons in initial
and final states and requiring that, the appropriate projections
do not change the theoretical expression for the three-point
correlation function, we may conclude, that in this limit there
are only two form-factors of order unity $G_1^V = G_1^A = \xi
(w)$, while all others are suppressed by heavy quark masses.

Having derived the spin symmetry relations, we came to situation
where we should calculate the only universal function in order to
obtain estimates on semileptonic or nonleptonic transitions of
doubly heavy baryons.

\subsection{\large\bf Spectral densities}

As, we said before, the calculation of spectral densities is
straightforward with the use of Cutkosky rules for the quark
propagators. However, the resulted expressions for NRQCD spectral
densities are different\footnote{It is simply an artifact of NRQCD
approximation. } for the cases of heavy to heavy or heavy to light
underlying quark transitions, so below we have classified the
calculated quantities. For the trace of correlation function with
$v^I_{\mu }$ we have

\vspace*{0.5cm} \noindent 1) heavy to heavy underlying transition

\begin{eqnarray}
\rho^{pert} &=& \frac{3m_1m_2}{\sqrt{\lambda
(s_I,s_F,q^2)}}(m_1^4-4m_1^2m_3^2+3m_3^4-
4m_1^3\sqrt{s_I}+8m_1m_3^2\sqrt{s_I}+\nonumber\\ &&
6m_1^2s_I-4m_3^2s_I -
4m_1s_I^{3/2}+s_I^2+4m_3^4\log\frac{\sqrt{s_I}-m_1}{m_3}) ,\\
\rho^{\bar qq} &=& -\frac{4 m_1 m_2 m_3
\sqrt{s_I}}{(m_1+m_3)\sqrt{\lambda (s_I,s_F,q^2)}}\langle\bar
qq\rangle ,
\end{eqnarray}

\noindent 2) heavy to light underlying transition

\begin{eqnarray}
\rho^{pert} &=& \frac{1}{4 (2\pi )^2}\frac{m_1}{\sqrt{s_I\lambda (s_I,s_F,q^2)}}
[-2(m_1-\sqrt{s_I})^6+3(m_1-\sqrt{s_I})^4(m_1^2+2m_3^2-\nonumber \\
&& q^2+2m_2\sqrt{s_I}+s_F)-6m_3^2(m_1-\sqrt{s_I})^2(2m_1^2+m_3^2-2q^2+4m_2\sqrt{s_I}+ \\
&& 2s_F)+m_3^4(9m_1^2+2m_3^2-9q^2+18m_2\sqrt{s_I}+9s_F)+12m_3^4(m_1^2-q^2+\nonumber \\
&& 2m_2\sqrt{s_I}+s_F)\log\frac{\sqrt{s_I}-m_1}{m_3}] , \nonumber \\
\rho^{\bar qq} &=& -\frac{m_1 m_3}{(m_1+m_3)\sqrt{\lambda (s_I,s_F,q^2)}}
(m_1^2-q^2+2m_2\sqrt{s_I}+s_F-m_3^2)\langle\bar qq\rangle ,
\end{eqnarray}
where
\begin{equation}
\lambda (x, y, z) = x^2 + y^2 + z^2 - 2xy - 2xz - 2yz.
\end{equation}
and the integration region in the double dispertion relation is
determined by the condition $$ -1 < \frac{1}{\sqrt{\lambda (s_I,
s_F, q^2)\lambda (s_I, m_1^2, m_3^2)}}
((s_I+s_F-q^2)(s_I+m_3^2-m_1^2)-2s_I(s_F-m_2^2+m_3^2)) < 1. $$ The
notations in the above expressions should be clear from Fig. 7.
Having derived theoretical expressions for the three-point
correlation function, we may proceed now with the evaluation of
form-factors. In numerical estimates we will use the Borel scheme
for the form-factor extraction and so, below we give an expression
determining the universal Isgur-Wise function for the semileptonic
decays of doubly heavy baryons
\begin{eqnarray}
\xi^{IW}(q^2) &=& \frac{1}{(2\pi )^2}\frac{1}{8 M_I M_F Z_I Z_F
}\int_{(m_1+m_3)^2}^{s_I^{th}}\int_{(m_1+m_2)^2}^{s_F^{th}}
\rho(s_I,s_F,q^2)ds_Ids_F\times \nonumber \\ && \exp
(-\frac{s_I-M_I^2}{B_I^2})\exp (-\frac{s_F-M_F^2}{B_F^2}),
\end{eqnarray}
where $B_I$ and $B_F$ are the Borel parameters in the initial and
final state channels.

\section{Numerical results}

In this section we give the results of numerical estimates on the
form-factors for the spin $1/2$ - spin $1/2$ doubly heavy baryon
transitions. Assuming the pole resonance model for the dependence
of mentioned form-factors on the square of lepton pair momentum we
make predictions on the semileptonic, pion and $\rho$ - meson
decay modes.

\subsection{Form-factors}

The analysis of NRQCD sum rules in the Borel scheme gives us the
estimates of the value of Isgur-Wise (IW) function at zero recoil
for different types of transitions between doubly heavy baryons,
shown in Table 1.

In Figs. 9-11 we have plotted the dependence of the normalization
of IW-function on the Borel parameters in the channels of initial
and final state baryons. Exploring the stability of NRQCD sum
rules upon variation of these parameters just give us the results
quoted above. The subtle point in the presented analysis is the
choice of the threshold values in the baryon channels. In the
present analysis we put the same values as in the analysis of
two-point sum rules. The results on the formfactors and later
their comparison with the results of potential models convince us,
that we made a right choice. However, the latter are in general
different from the ones used in two-point sum rule analysis in
moment scheme. For the situation, where it is the case we refer
the reader to \cite{Valera}.

\setlength{\unitlength}{1mm}
\begin{figure}[th]
\begin{center}
\vspace*{1.5cm}
\begin{picture}(200,150)
\put(50,-40){\epsfxsize=9cm \epsfbox{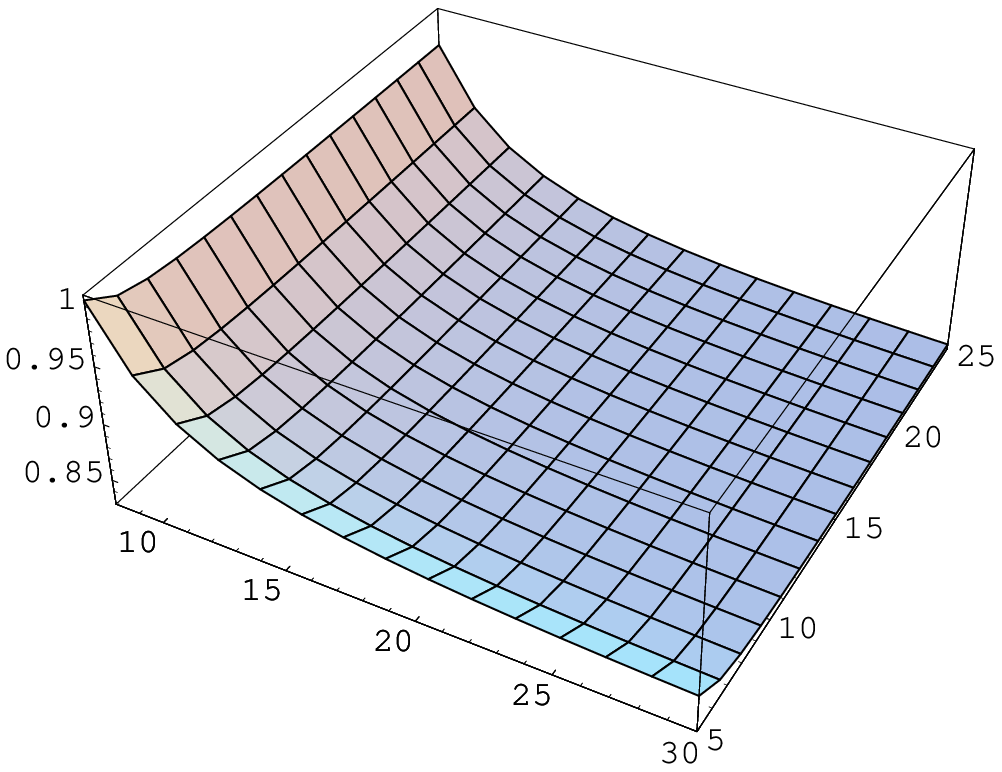}} \put(50,160){$\xi
(1)$} \put(75,115){$B_I$} \put(130,125){$B_F$}
\end{picture}
\end{center}
\vspace*{-10.cm} \caption{The value of $\xi (1)$ for the
transition $\Xi_{bb}^{\diamond}\to \Xi_{bc}^{\diamond}$ as
function of Borel parameters in the $s_I$ and $s_F$ channels.}
\label{3bb}
\end{figure}
\normalsize

\setlength{\unitlength}{1mm}
\begin{figure}[th]
\begin{center}
\vspace*{0.8cm}
\begin{picture}(200,150)
\put(50,-40){\epsfxsize=9cm \epsfbox{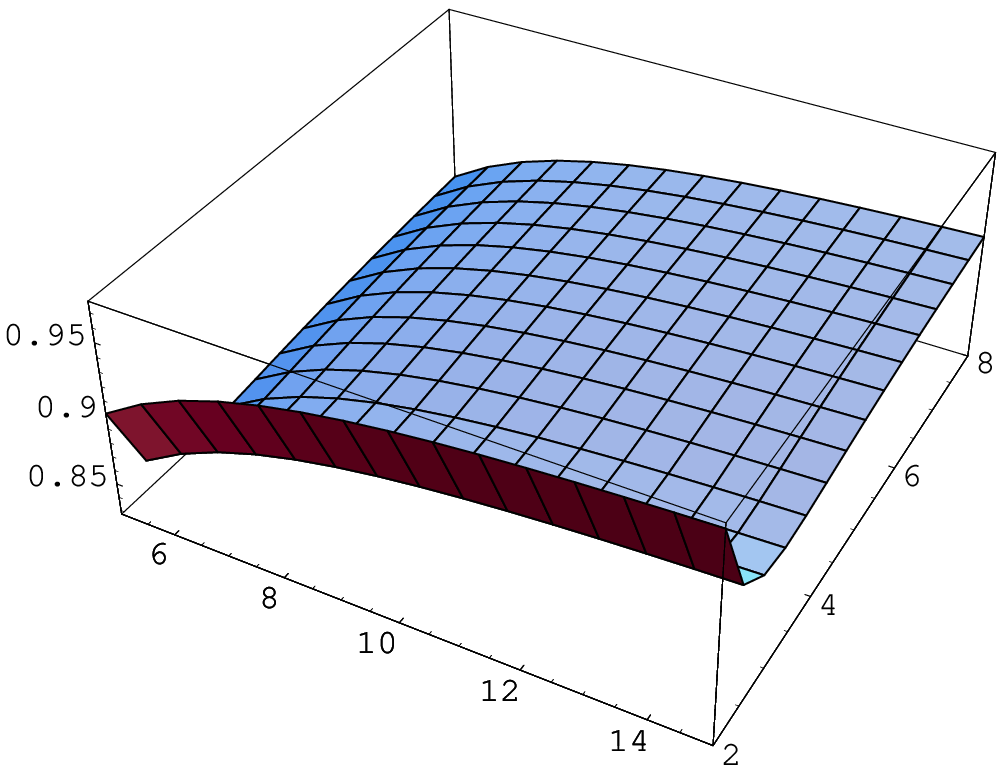}} \put(50,160){$\xi
(1)$} \put(75,115){$B_I$} \put(130,125){$B_F$}
\end{picture}
\end{center}
\vspace*{-10.cm} \caption{The value of $\xi (1)$ for the
transition $\Xi_{bc}^{\diamond}\to \Xi_{cc}^{\diamond}$ as
function of Borel parameters in the $s_I$ and $s_F$ channels.}
\label{3bc1}
\end{figure}
\normalsize

\setlength{\unitlength}{1mm}
\begin{figure}[th]
\begin{center}
\vspace*{0.5cm}
\begin{picture}(200,150)
\put(50,-40){\epsfxsize=9cm \epsfbox{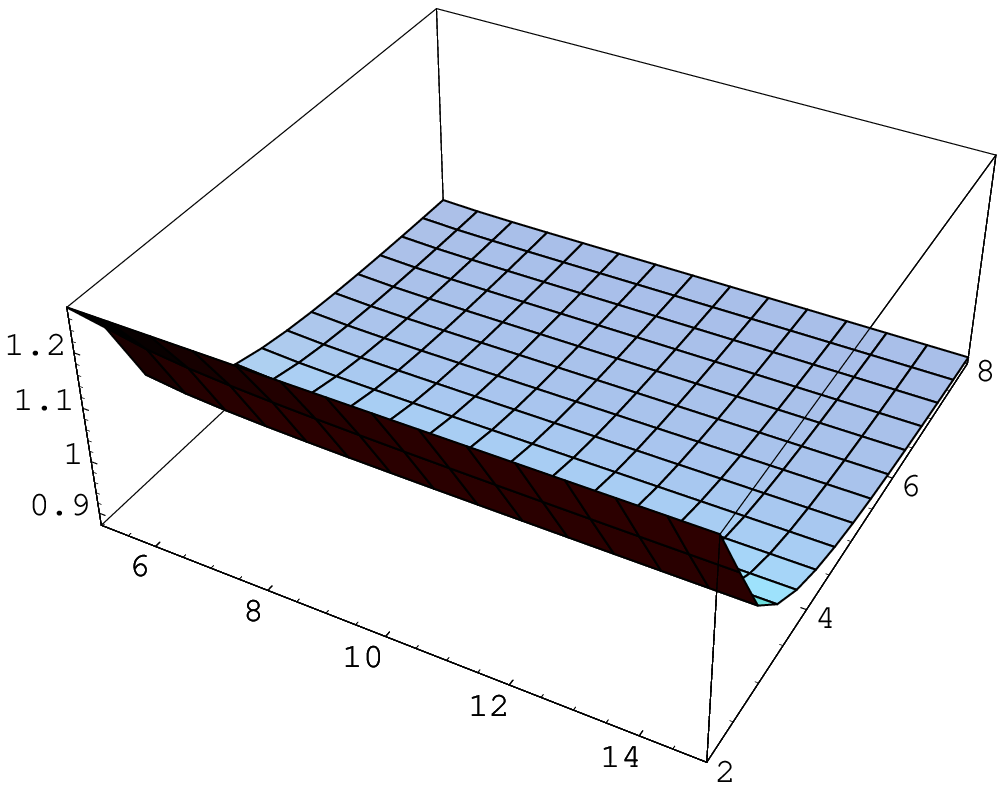}} \put(50,160){$\xi
(1)$} \put(75,115){$B_I$} \put(130,125){$B_F$}
\end{picture}
\end{center}
\vspace*{-10.cm} \caption{The value of $\xi (1)$ for the
transition $\Xi_{bc}^{\diamond}\to \Xi_{b}^{\diamond}$ as function
of Borel parameters in the $s_I$ and $s_F$ channels.} \label{3bc2}
\end{figure}
\normalsize

\setlength{\unitlength}{1mm}
\begin{figure}[th]
\begin{center}
\vspace*{0.5cm}
\begin{picture}(200,150)
\put(50,-40){\epsfxsize=9cm \epsfbox{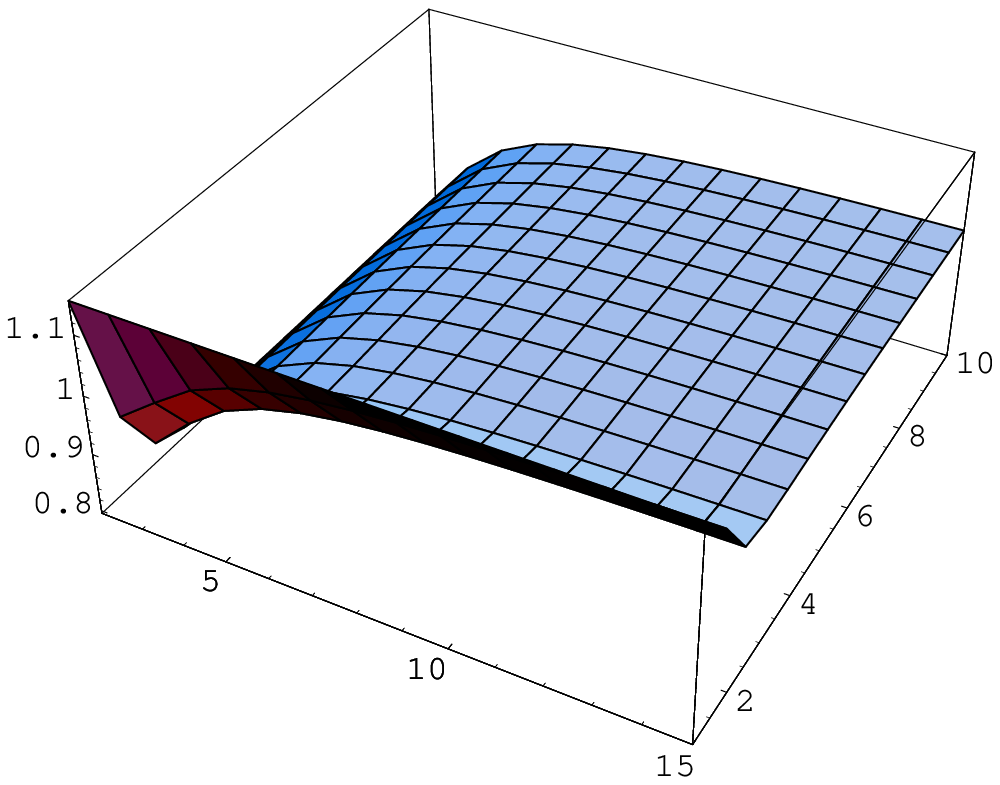}} \put(50,160){$\xi
(1)$} \put(75,115){$B_I$} \put(130,125){$B_F$}
\end{picture}
\end{center}
\vspace*{-10.cm} \caption{The value of $\xi (1)$ for the
transition $\Xi_{cc}^{\diamond}\to \Xi_{c}^{\diamond}$ as function
of Borel parameters in the $s_I$ and $s_F$ channels.} \label{3cc}
\end{figure}
\normalsize

For the sake of comparison, we also provide here the estimates of
the values of IW-function at zero recoil performed by us in the
framework of potential models (PM). In this approach the
normalization of IW-function is given by the overlap of initial
and final state baryon wave-functions. For simplicity we assume
the factorization of doubly heavy baryon wave-function in the
diquark and light quark - diquark wave-functions. From the
HQET-description of heavy mesons we know, that the light quark
affects the normalization of IW-function at zero recoil only in
$1/m_Q^2$ order. Thus, in our case its affect can be neglected and
the normalization of IW-function is given by the overlap of
diquark wave-functions. Taking the Gaussian anzaz for the diquark
wave-function  we get
\begin{eqnarray}
\xi^{IW}(1) = \left(\frac{2w_xw_y}{w_x^2+w_y^2}\right)^{3/2},
\end{eqnarray}
where
\begin{equation}
w_x = 2\pi\left(\frac{|Z_I|^2}{12}\right)^{1/3},\quad w_y =
2\pi\left(\frac{|Z_F|^2}{12}\right)^{1/3}
\end{equation}
and
\begin{equation}
|Z^{PM}| = 2\sqrt{3} |\Psi_d (0)\Psi_l (0)|.
\end{equation}
Here we have related the parameters of diquark wavefunctions to
the baryon couplings, obtained previously by us in the framework
of two-point NRQCD sum rules. In table 1 we have gathered the
results of sum rules and PM on the values of IW-functions at zero
recoil. We see, that within the errors of sum rules method (15 \%)
the obtained results are very close to those of PM.

\begin{table}[th]
\begin{center}
\begin{tabular}{|c|c|c|}
\hline\hline Mode & $\xi (1)$ SR &  $\xi (1)$ PM \\ \hline
$\Xi_{bb}\to \Xi_{bc}$ & 0.85 & 0.91 \\\hline $\Xi_{bc}\to
\Xi_{cc}$ & 0.91 & 0.99 \\\hline $\Xi_{bc}\to \Xi_{bs}$ & 0.9 &
0.99
\\\hline $\Xi_{cc}\to \Xi_{cs}$ & 0.99 & 1. \\\hline\hline
\end{tabular}
\end{center}
\caption{The normalization of Isgur-Wise function for different
baryon transitions at zero recoil.}
\end{table}

Next, to obtain the dependence of formfactors on the square of
momentum transfer we exploit the pole resonance model. So, for the
IW-function we have the following expression:
\begin{equation}
\xi^{IW} (q^2) = \xi_0\frac{1}{1-\frac{q^2}{m_{pole}^2}},
\end{equation}
with
\begin{eqnarray}
m_{pole} &=& 6.3 \mbox{~~~GeV for the~~}  b\to c
\mbox{~~transitions}\nonumber \\ m_{pole} &=& 1.85 \mbox{~~GeV for
the ~~} c\to s \mbox{~~transitions}.\nonumber
\end{eqnarray}

\subsection{Semileptonic decays}

Now knowing all formfactors, describing semileptonic transitions of doubly heavy baryons
we can estimate the semileptonic decay ratios for the transitions under consideration
\begin{eqnarray}
Br_{SL} (\Xi_{QQ'}^{\diamond}\to \Xi_{QQ'}^{\diamond '}) =
\tau_{\Xi_{QQ'}^{\diamond }} \int_1^{w_{max}}dw\frac{d\Gamma }{dw}
(\Xi_{QQ'}^{\diamond}\to \Xi_{QQ'}^{\diamond '}),
\end{eqnarray}
where
\begin{eqnarray}
w_{max} = \frac{M_I^2 + M_F^2 - m_l^2}{2M_IM_F};\quad q^2 = M_I^2
+ M_F^2 - 2M_IM_Fw.
\end{eqnarray}
For the $\frac{d\Gamma }{dw}$ we have
\begin{eqnarray}
\frac{d\Gamma }{dw} = \frac{d\Gamma_L}{dw} + \frac{d\Gamma_T}{dw},
\end{eqnarray}
where
\begin{eqnarray}
\frac{d\Gamma_L }{dw}(\Xi_{QQ'}^{\diamond}\to \Xi_{QQ'}^{\diamond
'}) &=& \frac{G_F^2}{(2\pi
)^3}|CKM|^2\frac{q^2M_F^2\sqrt{w^2-1}}{12M_I} \{ |H_{1/2,0}|^2 +
|H_{-1/2,0}|^2 \}, \\ \frac{d\Gamma_T
}{dw}(\Xi_{QQ'}^{\diamond}\to \Xi_{QQ'}^{\diamond '}) &=&
\frac{G_F^2}{(2\pi )^3}|CKM|^2\frac{q^2M_F^2\sqrt{w^2-1}}{12M_I}
\{ |H_{1/2,1}|^2 + |H_{-1/2,-1}|^2 \}.
\end{eqnarray}
Here $H_{\lambda_F,\lambda_W} = H_{\lambda_F,\lambda_W}^V -
H_{\lambda_F,\lambda_W}^A$, where $\lambda_F$ and $\lambda_W$ are
helicities of final state baryon and $W$ - boson correspondingly
and the functions $H_{\lambda_F,\lambda_W}^{V(A)}$ obey the
following symmetry relations:
\begin{equation}
H_{-\lambda_F,-\lambda_W}^{V(A)} = +(-)H_{\lambda_F,\lambda_W}^{V(A)}.
\end{equation}
The functions remained after the application of this relation can
be further expressed in terms of calculated in previous subsection
IW - functions with the help of the following formulae
\begin{eqnarray}
H_{1/2,1}^{V,A} &=& - 2\sqrt{M_IM_F(w\mp 1)}\xi^{IW}(w) \\
H_{1/2,0}^{V,A} &=&  \frac{1}{\sqrt{q^2}}\sqrt{2M_IM_F(w\mp
1)}(M_I\pm M_F)\xi^{IW}(w)
\end{eqnarray}
The results of numerical estimates, done with the help of
presented formulae can be found in Table 2.

To calculate the $\pi $ or $\rho $ - meson decays we assume the
hypothesis of factorization \cite{fact}. The corresponding
formulae for the decays of doubly heavy baryons with a pion or
$\rho $ - meson in the final state can be easily obtained from
those for the semileptonic decays by a simple substitution of
leptonic tensor by the $\pi $ - meson current tensor
$f_{\pi}^2p_{\mu}^{\pi}p_{\nu}^{\pi}$ or $\rho $ - meson current
tensor
$f_{\rho}^2m_{\rho}^2(-g_{\mu\nu}+p_{\mu}p_{\nu}/m_{\rho}^2)$
\begin{eqnarray}
\Gamma_{H_I\to H_F\pi} &=& 6\pi^2f_{\pi}^2a_1^2(\mu
)\frac{((M_I+M_F)^2-q^2)}{(M_I+M_F)^2}\frac{d\Gamma}{dq^2}
|_{q^2=m_{\pi}^2}, \\ \Gamma_{H_I\to H_F\rho } &=&
\frac{6\pi^2a_1^2(\mu )
f_{\rho}^2}{(M_I^2-M_F^2)^2}\{(M_I-M_F)^2((M_I+M_F)^2-q^2)+
\nonumber \\ && 2m_{\rho}^2((M_I-M_F)^2-q^2)\}\frac{d\Gamma
}{dq^2}|_{q^2=m_{\rho}^2},
\end{eqnarray}
where $a_1(\mu ) = \frac{1}{2N_c}(C_{+}(\mu )(N_c+1)+C_{-}(\mu
)(N_c-1))$ and $N_c=3$ is the number of colors. In numerical
calculations we put $a_1=1.26$.  The results for these nonleptonic
transitions can be also found in Table 2. To calculate the
branching ratios for exclusive decay modes we used the values of
doubly heavy baryon lifetimes, calculated by us previously
\cite{DHD}. There is some difference in concrete numerical values
of lifetimes, obtained in different papers \cite{ltime,DHD}. In
\cite{DHD} we have commented on the uncertainties in the resulting
values of lifetimes related to the heavy quark mass values. There
is, however, one more uncertainty remained, connected with the
value of light quark - diquark wave-function at origin. In present
there are two approaches to estimate this value: 1) assuming, that
this value is the same as the value of $D$-meson wave-function at
origin; 2) extracting this value from the comparison of hyper-fine
splittings in doubly and singly heavy baryons. Here we used the
estimates for the lifetimes made in the second approach, as they
are the most complete ones. The values, presented in Table 2
already include the contribution of spin $1/2$-spin $3/2$ decay
channels. To estimate the latter we have used the results of
\cite{Lozano}, where the contribution of these channels was
calculated for the case of $\Xi_{bc}\to \Xi_{cc}+l\bar\nu$ baryon
transition, and assumed, that, according to superflavor symmetry,
it constitutes 30 \% from the contribution of corresponding spin
$1/2$-spin $1/2$ transitions for all transitions between doubly
heavy baryons. In calculations of $\Xi_{bb}^{\diamond }$ and
$\Xi_{cc}^{\diamond }$ - baryon decay modes we have taken into
account a factor 2 due to Pauli principle for the identical heavy
quarks in the initial channel. In the case of $\Xi_{bc}^{\diamond
}\to \Xi_{cc}^{\diamond '} X$-baryon transition the same factor
comes from the positive Pauli interference of the $c$-quark, being
a product of $b$-quark decay, with the $c$-quark from the initial
baryon. Here, we also would like to mention, that for the
$\Xi_{bc}$-baryon decays the mentioned positive Pauli interference
contribution is dominant among the other nonspectator
contributions\footnote{Here we use the results of OPE analysis for
the inclusive decay modes of doubly heavy baryons
\cite{ltime,DHD}}, so we do not introduce other corrections here.
However, in the case of $\Xi_{cc}^{++}\to \Xi_{cs}^{+} X$- baryon
transition the negative Pauli interference plays the dominant role
and thus should be accounted for explicitly. From the previously
done OPE analysis for doubly heavy baryon lifetimes
\cite{ltime,DHD} we conclude that the corresponding correction
factor in this case is $0.62$. We would like also give a small
comment on our notations. The $\Xi_{Qs}^{\diamond}$ in Table 2
stays for the sum of $\Xi_{Q}^{\diamond}$ and $\Xi_{Q}^{\diamond
'}$ decay channels.

\begin{table}[th]
\begin{center}
\begin{tabular}{|c|c|c|c|}
\hline\hline Mode & Br (\%) &  Mode & Br (\%) \\ \hline
$\Xi_{bb}^{\diamond}\to \Xi_{bc}^{\diamond}l\bar\nu_l$ & 14.9 &
$\Xi_{bc}^{+}\to \Xi_{cc}^{++}l\bar\nu_l$ & 4.9 \\\hline
$\Xi_{bc}^{0}\to \Xi_{cc}^{+}l\bar\nu_l$ & 4.6 & $\Xi_{bc}^{+}\to
\Xi_{bs}^{0}\bar l\nu_l$ & 4.4 \\\hline $\Xi_{bc}^{0}\to
\Xi_{bs}^{-}\bar l\nu_l$ & 4.1 & $\Xi_{cc}^{++}\to
\Xi_{cs}^{+}\bar l\nu_l$ & 16.8 \\\hline $\Xi_{cc}^{+}\to
\Xi_{cs}^{0}\bar l\nu_l$ & 7.5 & $\Xi_{bb}^{\diamond}\to
\Xi_{bc}^{\diamond}\pi^{-}$ & 2.2
\\\hline $\Xi_{bb}^{\diamond}\to \Xi_{bc}^{\diamond}\rho^{-}$ &
5.7 & $\Xi_{bc}^{+}\to \Xi_{cc}^{++}\pi^{-}$ & 0.7 \\\hline
$\Xi_{bc}^{0}\to \Xi_{cc}^{+}\pi^{-}$ & 0.7 & $\Xi_{bc}^{+}\to
\Xi_{cc}^{++}\rho^{-}$ & 1.9 \\\hline $\Xi_{bc}^{0}\to
\Xi_{cc}^{+}\rho^{-}$ & 1.7 & $\Xi_{bc}^{+}\to
\Xi_{bs}^{0}\pi^{+}$ & 7.7 \\\hline $\Xi_{bc}^{0}\to
\Xi_{bs}^{-}\pi^{+}$ & 7.1 & $\Xi_{bc}^{+}\to
\Xi_{bs}^{0}\rho^{+}$ & 21.7 \\\hline $\Xi_{bc}^{0}\to
\Xi_{bs}^{-}\rho^{+}$ & 20.1 & $\Xi_{cc}^{++}\to
\Xi_{cs}^{+}\pi^{+}$ & 15.7 \\\hline $\Xi_{cc}^{+}\to
\Xi_{cs}^{0}\pi^{+}$ & 11.2 & $\Xi_{cc}^{++}\to
\Xi_{cs}^{+}\rho^{+}$ & 46.8 \\\hline $\Xi_{cc}^{+}\to
\Xi_{cs}^{0}\rho^{+}$ & 33.6 &  & \\\hline\hline
\end{tabular}
\end{center}
\caption{Branching ratios for the different decay modes of doubly
heavy baryons. }
\end{table}

The previous studies of exclusive decays of doubly heavy baryons
\cite{Lozano,Guo} exploited the spin-flavor symmetry of QCD
\cite{SS}, arising at the very large quark mass limit. The
fundamental representation of the SU(6)$\otimes $U(1) spin-flavor
symmetry group consists of two-component heavy quark spinor,
scalar and axial-vector di-antiquark fields. This representation
can be explicitly written in terms of nine-component vector with
the four-velocity $v$:
\begin{center}
\[ {\Psi}_v\ =\ \left( \begin{array}{c}
h_v \\ S_v \\ A_v^{\mu} \end{array} \right) \]
\end{center}
where $A_v^{\mu} $ satisfies the constraint $v_{\mu}A_v^{\mu} = 0$
and the effective lagrangian for this field is
\[ {\cal L}_{eff}\ =\ \frac{1}{2}\ \overline{\Psi}_v{\cal M}\ iv{\cdot}
^{^<}\overline{D}^{>}\ {\Psi}_v \] where $\cal M$ is a 9$\times $9
mass matrix\footnote{Note, that the particular form of the mass
matrix depends on the fields normalizations.}, given by the
following expression:
\[ {\cal M}\ =\ \left(
\begin{array}{ccc}
1 & 0 & 0 \\ 0 & 2m_S & 0 \\ 0 & 0 & -2m_A
\end{array} \right) \]
and
\[ \overline{\Psi}_v\ =\ {\Psi}_v^{\dag}\ \left(
\begin{array}{rrr}
{\gamma}^0\ 0\ 0 \\ 0\ 1\ 0 \\ 0\ 0\ g
\end{array} \right) \]
Here $g = \mbox{diag(1,-1,-1,-1)}$ is the usual metric tensor.
Next, to make connection with the hadronic states, one considers a
tensor product of $\Psi_v$ with one light antiquark field. Thus,
this hadronic supermultiplet puts together singly heavy mesons and
doubly heavy antibaryons. However, such supermultiplet is not
completely flavor-independent even in the heavy quark mass limit,
as there remains internal mass-dependent heavy di-antiquark
dynamics. The singly heavy baryons with the strangeness in the
discussed approach belong to the different supermultiplet and this
fact should be taken into account, when calculating form-factors
for the semileptonic transitions between doubly heavy and singly
heavy baryons. Such analysis within the framework of potential
models was performed previously by M.A.Sanchis-Lozano
\cite{Lozano} for the case of $\Xi_{bc}^{\diamond }$-baryon
decays. There, to calculate the form-factors, it was assumed that
the latter are given by the overlap of Coulomb diquark
wave-functions with small non-perturbative corrections, given by
the presence of a light quark in the baryons under consideration.
It is just the approach we have used in our PM estimates. To
reduce the number of independent form-factors, there was performed
an analysis of spin-symmetry relations between various
form-factors in the limit of zero recoil. The author, using
different arguments, had came to the same conclusion as we have
did in the present work. That is, all semileptonic transitions of
doubly heavy baryons are governed by the only universal function,
an analogue of Isgur-Wise function. The numerical results on the
normalization of Isgur-Wise function at zero recoil completely
agree with our estimates both in the framework of potential models
and NRQCD sum rules. The given predictions for the semileptonic
decay modes of $\Xi_{bb}^{\diamond}$-baryons  nicely agree with
the ones presented in this paper, taking into account correction
factor due to the wrong values of $\Xi_{bc}^{\diamond}$-baryon
lifetime used in that paper. There is also a paper, where the
diquark semileptonic transitions where calculated  within the
Bethe-Salpeter approach. However, the numerical results presented
in this paper are very strange. It is suffice to say, that, for
example, according to these results the semileptonic branching
ratios of $\Xi_{bb}^{\diamond }$-baryon decays should be
approximately 50 \%, what is very unlikely.

To finish the discussion of the obtained results we would like to
note, that the latter are also in agreement with the estimates of
inclusive decay channels performed by us previously
\cite{ltime,DHD}.

\section{Conclusion}

In this paper we have presented the analysis of exclusive decays
of doubly heavy baryons in the framework of NRQCD sum rules. We
have provided  complete numerical study of baryonic couplings and
semileptonic form-factors. The values of semileptonic and some
nonleptonic exclusive modes are also given. To conclude, we would
like to discuss what also can be and should be done in the study
of exclusive decays of doubly heavy baryons. First, it will be
instructive to perform the similar analysis for the baryon
currents of the first type. We have check, that these two schemes
of calculation give the similar results only in the case of
$\Xi_{bb}^{\diamond }$-baryon decays. Second, one may perform an
analysis of doubly heavy baryon exclusive decays in full QCD and
not relay on the pole resonance model for the form-factors. We
plan to present the results of such analysis in nearest future.
And, third one may try to calculate the lifetimes of doubly heavy
baryons in the framework of QCD sum rules. It is a very
interesting task, as we will explicitly see the effect of large
nonspectator effects, studied previously in the OPE framework, on
various exclusive modes.

The author is grateful to Prof. V.V.Kiselev and A.E.Kovalsky for
reading this manuscript and making valuable remarks. I especially
thank my wife for strong moral support and help in doing physics.

This work was in part supported by the Russian Foundation for
Basic Research, grants 99-02-16558 and 00-15-96645, by
International Center of Fundamental Physics in Moscow,
International Science Foundation and INTAS-RFBR-95I1300.

\newpage

\appendix
\section*{Appendix A}

In this Appendix we have collected theoretical expressions for spectral
densities of Wilson coefficients, standing in front of various operators, obtained
as the result of OPE expansion of two-point correlation function.

For the case of $\epsilon^{\alpha\beta\lambda }:(Q_{\alpha }^{'T}C\gamma_5
q_{\beta})Q_{\lambda }:$ current we have
\begin{eqnarray}
\rho_1^{pert}   &=& \frac{2\sqrt{2}\sqrt{m_1m_2(m_1+m_2)}}{105(m_1+m_2)^3\pi^3}w^{7/2}
(m_1m_2(12m_2-13w)+5m_2^2w+m_1^2(12m_2+5w))\nonumber    \\
&& \\
\rho_2^{pert} &=& \frac{2\sqrt{2}\sqrt{m_1m_2(m_1+m_2)}}{105(m_1+m_2)^2\pi^3}w^{7/2}
(m_1m_2(12m_2-w)+m_2^2w+m_1^2(12m_2+5w)) \\
\rho_1^{\bar qq} &=& -\frac{\sqrt{m_1m_2(m_1+m_2)}}{4\sqrt{2}(m_1+m_2)^3\pi}\sqrt{w}
(m_1m_2(4m_2-5w)+5m_2^2w+m_1^2(4m_2+w)) \\
\rho_2^{\bar qq} &=& -\frac{\sqrt{m_1m_2(m_1+m_2)}}{4\sqrt{2}(m_1+m_2)^2\pi}\sqrt{w}
(m_1m_2(4m_2-w)+m_2^2w+m_1^2(4m_2+w)) \\
\rho_1^{G^2} &=& \frac{1}{1536\sqrt{2}\pi (m_1m_2)^{3/2}(m_1+m_2)^{7/2}}\sqrt{w}
(2 m_2^5w^2-28m_1^4m_2w(4m_2+w)+\nonumber \\
&& m_1m_2^4w(-16m_2+35w)+m_1^5(32m_2^2+8m_2w-w^2) - \\
&& 8m_1^2m_2^3(8m_2^2-13m_2w+28w^2)+m_1^3(-96m_2^4+217m_2^2w^2))\nonumber \\
\rho_2^{G^2} &=& \frac{1}{7680\sqrt{2}\pi (m_1m_2)^{3/2}(m_1+m_2)^{5/2}}\sqrt{w}
(122m_2^5w^2-100m_1^4m_2(4m_2+w)+\nonumber \\
&& m_1m_2^4w(880m_2+327w)-40m_1^2m_2^3(8m_2^2-41m_2w-20w^2)+ \\
&& 5m_1^2(32m_2^2+8m_2w-w^2)+5m_1^3m_2^2(-96m_2^2+64m_2w+169w^2))\nonumber \\
\rho_1^{mix} &=& \frac{1}{2048\sqrt{2}\pi (m_1m_2)^{1/2}(m_1+m_2)^{7/2}\sqrt{w}}
(m_1^2m_2^2(64m_2-397w)+ \\
&& 30m_1m_2^3(4m_2-27w)+105m_2^4w+m_1^4(-104m_2+17w)+10m_1^3m_2(-16m_2+19w))\nonumber \\
\rho_2^{mix} &=& \frac{1}{2048\sqrt{2}\pi (m_1m_2)^{1/2}(m_1+m_2)^{5/2}\sqrt{w}}
(2m_1^3m_2(96m_2-43w)- \\
&& 6m_1m_2^3(4m_2-23w)+15m_2^4w+m_1^4(104m_2-17w)+m_1^2m_2^2(64m_2+45w))\nonumber
\end{eqnarray}
Here $m_1$ is the mass of $Q$-quark and $m_2$ is the mass of
$Q'$-quark. For the case of $\epsilon^{\alpha\beta\lambda
}:(Q_{\alpha }^{T}C\gamma_5 q_{\beta})s_{\lambda }:$ current we
have
\begin{eqnarray}
\rho_1^{pert} &=& \frac{w^3}{80m_Q^2\pi^3 }(135m_s^2w^2-12m_Qm_sw(5m_s+3w)+
4m_Q^2(5m_s^2+5m_sw+w^2)) \\
\rho_2^{pert} &=& \frac{m_sw^3}{40m_Q\pi^3 }(9m_sw^2+5m_Q^2(4m_s+w)-m_Qw(5m_s+w)) \\
\rho_1^{\bar qq} &=& -\frac{1}{4m_Q^3\pi }(-85m_s^2w^3-m_Q^2w(3m_s+2w)^2+
m_Qm_sw^2(33m_s+26w)+\nonumber\\
&& m_Q^3(ms^2+4m_sw+2w^2)) \\
\rho_2^{\bar qq} &=& -\frac{m_Qm_s}{4\pi (m_Q+w)^2}(2m_Q^2(m_s+w)+w^2(m_s+w)+m_Qw(2m_s+3w)) \\
\rho_1^{G^2} &=& \frac{1}{1536\pi m_Q^4}(-2185m_s^2w^3-28m_Q^2w(3m_s+2w)^2+
2m_Qm_sw^2(437m_s+344w)-\nonumber \\
&& 32m_Q^4(m_s+w)+32m_Q^3(m_s^2+4m_sw+2w^2))   \\
\rho_2^{G^2} &=& -\frac{1}{384\pi m_Q^2}(m_Qm_sw(6m_s-13w)+4m_s^2w^2+8m_Q^2m_s(w(\log 8-4)+
m_s(\log 8-1))-\nonumber \\
&& 8m_Q^3(m_s+4w-m_s\log 8)+24m_Q^2m_s(m_Q+m_s+w)\log w) \\
\rho_1^{\bar qGq} &=& \frac{1}{16m_Q^4\pi}(m_Q^4+168m_s^2w^2-5m_Q^3(m_s+w)-
3m_Qm_sw(18m_s+19w)+\nonumber \\
&& m_Q^2(10m_s^2+22m_sw+11w^2)) \\
\rho_2^{\bar qGq} &=& \frac{m_s}{16m_Q^3\pi}(-m_Q^3+3m_sw^2+m_Q^2(m_s+w)-m_Qw(2m_s+w))  \\
\rho_2^{\bar sGs} &=& \frac{m_Q+m_s+w}{16\pi}
\end{eqnarray}


\begin{thebibliography}{**}
\bibitem{cdf-bc}
{ F. Abe et al.}, CDF Collaboration, Phys. Rev. Lett. {\bf 81},
2432 (1998), Phys. Rev. {\bf D58}, 112004 (1998).
\bibitem{bc-rev}
{ S.S.Gershtein, V.V.Kiselev, A.K.Likhoded, A.V.Tkabladze,
A.V.Berezh\-noy, A.I.Onish\-chen\-ko}, Talk given at 4th
International Workshop on Progress in Heavy Quark Physics,
Rostock, Germany, 20-22 Sept. 1997, IHEP 98-22 [hep-ph/9803433];\\
{ S.S.Gershtein, V.V.Kiselev, A.K.Likhoded, A.V.Tkabladze}, Phys.
Usp. {\bf 38}, 1 (1995) [Usp. Fiz. Nauk {\bf 165}, 3 (1995)];\\ {
S.S.Gershtein et al.}, Phys. Rev. {\bf D51}, 3613 (1995).
\bibitem{prod}
{ A.V.Berezhnoy, V.V.Kiselev, A.K.Likhoded, A.I.Onishchenko},
Phys. Rev. D57 (1998) 4385;\\ { A.V.Berezhnoy, V.V.Kiselev,
A.K.Likhoded}, Z.Phys. {\bf A356}, 89 (1996), Phys. Atom. Nucl.
{\bf 59}, 870 (1996) [Yad. Fiz. {\bf 59}, 909 (1996)];\\ {
S.P.Baranov}, Phys. Rev. D 56, 3046 (1997);\\ { V.V.Kiselev, A.K.
Likhoded, M.V. Shevlyagin}, Phys. Lett. B332, 411 (1994);\\ {
A.Falk et al.}, Phys. Rev. D49, 555 (1994);\\ V.V.Kiselev,
A.E.Kovalsky, preprint hep-ph/9908321.
\bibitem{ltime}
{ V.V.Kiselev, A.K.Likhoded, A.I.Onishchenko}, Phys. Rev. {\bf
D60}, 014007 (1999), Phys.Atom.Nucl. {\bf 62} (1999) 1940,
Yad.Fiz. {\bf 62} (1999) 2095; \\ { V.V.Kiselev, A.K.Likhoded,
A.I.Onishchenko}, preprint DESY 98-212 (1999) [hep-ph/ 9901224],
to appear in Eur.Phys.J. {\bf C} ;\\ { B.Guberina, B.Melic,
H.Stefancic}, Eur. Phys. J. {\bf C9}, 213 (1999).
\bibitem{DHD}
{A.I.Onishchenko}, preprint hep-ph/9912424; \\ {A.K.Likhoded,
A.I.Onishchenko}, preprint hep-ph/9912425.
\bibitem{pot}
{ S.S.Gershtein, V.V.Kiselev, A.K.Likhoded, A.I.Onishchenko},
preprint IHEP 98-66 (1998) [hep-ph/9811212], Heavy Ion Phys {\bf
9}, 133 (1999); [hep-ph/9807375] Mod. Phys. Lett. {\bf A14}, 135
(1999);\\ { D.Ebert, R.N.Faustov, V.O.Galkin, A.P.Martynenko,
V.A.Saleev}, Z. Phys. C76 (1997) 111;\\ { J.G.K$\ddot o$rner,
M.Kr$\ddot a$mer, D.Pirjol}, Prog. Part. Nucl. Phys. 33 (1994)
787;\\ { R.Roncaglia, D.B.Lichtenberg, E.Predazzi}, Phys. Rev. D52
(1995) 1722.
\bibitem{SVZ}
{ M.A.Shifman, A.I.Vainshtein, V.I.Zakharov}, Nucl. Phys. {\bf
B147}, 385 (1979);\\ { L.J.Reinders, H.R.Rubinstein, S.Yazaki},
Phys. Rep. {\bf 127}, 1 (1985).
\bibitem{QCDsr}
{ E.Bagan, M.Chabab, S.Narison}, Phys. Lett. B306 (1993) 350;\\ {
E.Bagan at al.}, Z. Phys. {\bf C64}, 57 (1994).
\bibitem{DHSR1}
{V.V.Kiselev, A.I.Onishchenko}, preprint hep-ph/9909337, to appear
in Nucl.Phys. {\bf B}.
\bibitem{DHSR2}
{V.V.Kiselev, A.E.Kovalsky}, preprint hep-ph/0005019.
\bibitem{Lozano}
{M.A.Sanchis-Lozano}, Nucl.Phys. {\bf B440} (1995) 251.
\bibitem{Guo}
{X.-H.Guo, H.-Y.Jin, X.-Q.Li}, Phys.Rev {\bf D58} (1998), 114007.
\bibitem{Smilga}
{ A.Smilga}, Yad. Phys. {\bf 35}, 473 (1982).
\bibitem{Cutk}
{ R.E.Cutkosky}, J. Math. Phys. {\bf 1}, 429 (1960).
\bibitem{Valera}
{ V.V.Kiselev, A.V.Tkabladze}, Phys. Rev. {\bf D48},(1993), 5208;
\\ { V.V.Kiselev, A.K.Likhoded, A.I.Onishchenko}, Nucl.Phys.
{\bf B569}, (2000), 473; { V.V.Kiselev, A.E.Kovalsky,
A.K.Likhoded}, hep-ph/0002127
\bibitem{fact}
M.Dugan and B.Grinstein, Phys.Lett. {\bf B255} (1991) 583;\\
M.A.Shifman, Nucl.Phys. {\bf B388} (1992) 346; \\ B.Blok,
M.Shifman, Nucl.Phys. {\bf 389} (1993) 534. \bibitem{SS} H.Georgi
and M.B.Wise, Phys.Lett. {\bf B243} (1990) 279;\\ C.D.Carone,
Phys.Lett. {\bf B253} (1991) 408;\\ M.J.Savage and M.B.Wise,
Phys.Lett. {\bf B248} (1990) 177.
\end{thebibliography}
\end{document}